\UseRawInputEncoding
\documentclass[referee]{aa} 

\usepackage[utf8]{inputenc}
\usepackage{graphicx}
\usepackage{txfonts}
\usepackage{hyperref}
\begin{document}

   \title {FRIPON: A worldwide network to track incoming meteoroids}
   



 \author{
F. Colas                \inst{1,9}
\thanks{Corresponding authors: Francois.Colas@obspm.fr, Brigitte.Zanda@mnhn.fr, Mirel.Birlan@obspm.fr}
                 \and     
B. Zanda                \inst{2,1,9}\and   
S. Bouley               \inst{3,1,9}\and   
S. Jeanne               \inst{1,9}\and     
A. Malgoyre             \inst{7,9}\and     
M. Birlan               \inst{1,9}\and     
C. Blanpain             \inst{7,9}\and     
J. Gattacceca   \inst{5,9}\and     
L. Jorda                \inst{4,9}\and      
J. Lecubin              \inst{7,9}
% ...
 }

\institute{
$^{1}$  IMCCE, Observatoire de Paris, PSL Research University, CNRS UMR 8028, Sorbonne Université, Université de Lille, 77 av. Denfert-Rochereau, 75014, Paris, France.\\   
$^{2}$  Institut de Minéralogie, Physique des Matériaux et Cosmochimie (IMPMC), Muséum National d'Histoire Naturelle, CNRS UMR 7590, Sorbonne Université, F-75005 Paris, France.\\   
$^{3}$  GEOPS-Géosciences, CNRS, Université Paris-Saclay, 91405, Orsay, France.\\ 
$^{4}$  Aix Marseille Univ, CNRS, CNES, LAM, Marseille, France.\\
$^{5}$  Aix Marseille Univ, CNRS, IRD, Coll France, INRA, CEREGE, Aix-enProvence, France.\\  
$^{6}$  Aix Marseille Universit\'e, CNRS, OSU Institut Pyth\'eas UMR 3470 Marseille, France.\\
$^{7}$  Service Informatique Pythéas (SIP) CNRS - OSU Institut Pythéas - UMS 3470, Marseille, France.\\   
$^{8}$  IAS, CNRS, Université Paris-Saclay, 91405, Orsay, France.\\   
$^{9}$  FRIPON (Fireball Recovery and InterPlanetary Observation) and Vigie-Ciel Team, France.\\ 
$^{10}$ INAF - Osservatorio Astrofisico di Torino -     Via Osservatorio 20, 10025 Pino Torinese, TO, Italy.\\  
$^{11}$ Università degli Studi di Torino,  Dipartimento di Fisica,     Via Pietro Giuria 1, 10125 Torino, TO, Italy.\\
$^{12}$ INAF - Osservatorio di Astrofisica e Scienza dello Spazio       Via Piero Gobetti 93/3, 40129 Bologna, BO, Italy.\\
$^{13}$ INAF - Osservatorio Astronomico di Trieste      Via Giambattista Tiepolo 11, 10134 Trieste, TS, Italy.\\
$^{14}$ Università degli Studi di Firenze - Dipartimento di Scienze della Terra   Via Giorgio La Pira, 4, 50121 Firenze, FI, Italy.\\
$^{15}$ Osservatorio Astronomico del Righi      Via Mura delle Chiappe 44R, 16136 Genova, GE, Italy.\\ 
$^{16}$ Planétarium Rio Tinto Alcan / Espace pour la vie, Montréal, Québec, Canada.\\
$^{17}$ Réseau DOME, (Détection et Observation de Météores / Detection and Observation of Meteors), Canada.\\
$^{18}$ Division for Medical Radiation Physics and Space Environment, University of Oldenburg, Germany.\\
$^{19}$ Dep. Física Aplicada I, Escuela de Ingeniería de Bilbao, Universidad del País Vasco/Euskal Herriko Unibertsitatea, 48013 Bilbao, Spain.\\
$^{20}$ Aula EspaZio Gela, Escuela de Ingeniería de Bilbao, Universidad del País Vasco/Euskal Herriko Unibertsitatea, 48013 Bilbao, Spain.\\
$^{21}$ Università degli Studi di Firenze - Museo di Storia Naturale   Via Giorgio La Pira, 4, 50121 Firenze, FI, Italy.\\ 
$^{22}$ INAF - Istituto di Astrofisica e Planetologia Spaziali Via del Fosso del Cavaliere 100, 00133 Roma, RM, Italy.\\
$^{23}$ CNR - Istituto di Fisica Applicata Nello Carrara        Via Madonna del Piano, 10 50019 Sesto Fiorentino (FI), Italy.\\
$^{24}$ INAF - Osservatorio Astronomico di Padova       Vicolo dell'Osservatorio 5, 35122 Padova, PD, Italy.\\
$^{25}$ INAF - Istituto di Radioastronomia      Via Piero Gobetti 101, 40129 Bologna, BO, Italy.\\
$^{26}$ INAF - Osservatorio Astronomico d'Abruzzo       Via Mentore Maggini snc, Loc. Collurania, 64100 Teramo, TE, Italy.\\
$^{27}$ INAF - Osservatorio Astrofisico di Catania      Via Santa Sofia 78, 95123 Catania, CT, Italy.\\
$^{28}$ INAF - Osservatorio Astrofisico di Arcetri      Largo Enrico Fermi 5, 50125 Firenze, FI, Italy.\\
$^{29}$ INAF - Osservatorio Astronomico di Cagliari     Via della Scienza 5, 09047 Cuccuru Angius, Selargius, CA, Italy.\\
$^{30}$ INAF - Osservatorio Astronomico di Brera        Via Brera 28, 20121 Milano, MI, Italy.\\
$^{31}$ FRIPON-Chile.\\ 
$^{32}$ Natural History Museum, Burgring 7, A-1010 Vienna, Austria.\\ 
$^{33}$ FRIPON-Belgium.\\ 
$^{34}$ SPMN (SPanish Meteor Network), FRIPON, Spain.\\ 
$^{35}$ FRIPON-Netherlands, European Space Agency, SCI-SC, Keplerlaan 1, 2201 AZ Noordwijk, Netherlands.\\ 
$^{36}$ PRISMA (Prima Rete per la Sorveglianza sistematica di Meteore e Atmosfera), Italy.\\ 
$^{37}$ MOROI (Meteorites Orbits Reconstruction by Optical Imaging) Astronomical Institute of the Romanian Academy, Bucharest, Romania.\\ 
$^{38}$ Oukaimeden Observatory, High Energy Physics and Astrophysics Laboratory, Cadi Ayyad University, Marrakech, Morocco.\\ 
$^{39}$ BRAMON (Brazilian Meteor Observation Network), Brazil.\\ 
$^{40}$ FRIPON-Germany.\\ 
$^{41}$ FRIPON-Switzerland.\\ 
$^{42}$ Università di Padova - Dipartimento di Fisica e Astronomia     Vicolo dell'Osservatorio 3, 35122 Padova, PD, Italy.\\
$^{43}$ Universciences, 30 avenue Corentin Cariou, 75019 Paris, France.\\
$^{44}$ REFORME (REseau Français d'ObseRvation de MEtéore) France.\\
$^{45}$ SCAMP (System for Capture of Asteroid and Meteorite Paths), FRIPON, UK.\\ 
$^{46}$ Natural History Museum,Cromwell Road, London, UK.\\
$^{47}$ Cosmos Sterrenwacht, 7635 NK Lattrop, Netherlands.\\
$^{48}$ Cyclops Observatory, 4356 CE Oostkapelle, Netherlands.\\
$^{49}$ KVI - Center for Advanced Radiation Technology, Zernikelaan 25, 9747 AA Groningen, Netherlands.\\
$^{50}$ Leiden Observatory, 2333 CA Leiden, Netherlands.\\
$^{51}$ European Space Agency, OPS-SP, Keplerlaan 1, 2201 AZ Noordwijk, Netherlands.\\
$^{53}$ International Meteor Organization.\\   
$^{54}$ Espace des Sciences, Planétarium, Rennes, France.\\
$^{55}$ Université de technologie de Compiègne,  (Multi-scale modeling of urban systems), Centre Pierre Guillaumat - Université de Technologie de Compiègne, 60200 Compiègne, France.\\
$^{56}$ Lycée Saint-Paul, 12 allée Gabriel Deshayes, 56017 Vannes, France.\\
$^{57}$ Station de Radioastronomie de Nançay, 18330 Nançay, France.\\
$^{58}$ GISFI, Rue Nicolas Copernic, 54310 Homécourt, France.\\
$^{59}$ Geosciences Environnement Toulouse, UMR5563 CNRS, IRD et Université de Toulouse, 14 avenue Edouard Belin,31400 Toulouse, France.\\
$^{60}$ FRIPON, Algeria.\\
$^{61}$ Cerap – Planétarium de Belfort, Cité des associations 90000 Belfort, France.\\
$^{62}$ Club d'Astronomie du FLEP - "La rampisolle" 24660 Coulounieix-Chamiers, France.\\
$^{63}$ Les Editions du Piat, Glavenas, 43200 Saint-Julien-du-Pinet, France.\\
$^{64}$ Université Grenoble Alpes, CNRS, IPAG, 38400 Saint-Martin d’Hères, France.\\
$^{65}$ Institut Universitaire de France, Paris, France. \\
$^{66}$ Geneva Observatory, CH-1290 Sauverny, Switzerland.\\
$^{67}$ PALEVOPRIM (Laboratoire Paléontologie Evolution Paléo Écosystèmes Paléoprimatologie), (iPHEP, UMR-CNRS 7262), UFR SFA,Université de Poitiers, 86022 Poitiers, France .\\
$^{68}$ LPG-BIAF Faculté des sciences Géologie 49045 - Poitiers France.\\
$^{69}$ FRIPON-Morocco.\\
$^{70}$ Observatoire Astronomique de Valcourt, 52100 Valcourt France.\\
$^{71}$ Planétarium LUDIVER, 1700, rue de la libération Tonneville 50460 La Hague, France.\\
$^{72}$ Association Astronomique de Belle-Ile-en-mer 56360 Bangor, France.\\
$^{73}$ Le Planétarium Roannais 42153 Riorges, France.\\
$^{74}$ Bucharest University, Faculty of Physics, 405 Atomistilor, 077125 Magurele, Ilfov, Romania.\\
$^{75}$ Groupe Astronomique de Querqueville, 50460 Cherbourg en Cotentin, France.\\
$^{76}$ Laboratoire Interdisciplinaire Carnot de Bourgogne, UMR 6303 CNRS/Univ. Bourgogne Franche-Comté Dijon, France.\\
$^{77}$ Société astronomique du Haut Rhin - 68570 Osenbach, France.\\
$^{78}$ European Southern Observatory, Alonso de Córdova 3107, Vitacura, Santiago, Chile.\\
$^{79}$ Observatoire de Gramat, 46500 Gramat, France.\\
$^{80}$ Carrefour des Sciences et des Arts, 46000 Cahors, France.\\
$^{81}$ Université Clermont Auvergne, CNRS, IRD, OPGC, Laboratoire Magmas et Volcans, F-63000 Clermont-Ferrand, France.\\
$^{82}$ INU champollion dphe, Place de verdun, 81000 Albi, France.\\
$^{83}$ Observatório Nacional/MCTI, R. General José Cristino 77, Rio de Janeiro – RJ 20921-400, Brazil.\\ 
$^{84}$ Stella Mare - Universta di Corsica - CNRS - 20620 Biguglia, France.\\
$^{85}$ Association Astronomique "Les têtes en l'air", Marigny, France.\\
$^{86}$ Pôle des étoiles, Route de Souesmes, 18330 Nançay, France.\\
$^{87}$ LPC2E, University of Orleans, CNRS, Orléans, France.\\
$^{88}$ FRIPON - Per\'u.\\
$^{89}$ CRPG - CNRS, 15 Rue Notre Dame des Pauvres, 54500 Vand \oe{}uvre-lès-Nancy, France.\\
$^{90}$ Observatoire de la Lèbe, Chemin des étoiles, 01260 Valromey-sur-Séran, France.\\
$^{91}$ Armagh Observatory and Planetarium, Armagh, Northern Ireland, UK.\\
$^{92}$ Laboratoire Géosciences Appliquées à l’ingénierie de l'Aménagement GAIA - Université Hassan II de Casablanca, Faculté des Sciences Ain Chock, Casablanca, Marocco.\\
$^{93}$ Shelyak Instruments, 77 Rue de Chartreuse, 38420 Le Versoud, France.\\
$^{94}$ Parc du Cosmos, 30133 Les Angles, France.\\
$^{95}$ Écomusée de la Baie du Mont Saint-Michel, 50300 Vains Saint-Léonard, France.\\
$^{96}$ Association Science en Aveyron, 12000 Rodez, France.\\
$^{97}$ CNRS, Aix-Marseille Université, PIIM UMR 7345, Marseille, France. \\
$^{98}$ Observatoire de Narbonne, 11100 Narbonne, France. \\
$^{99}$ Muséum des Volcans 15000 Aurillac, France.\\
$^{100}$ Académie des sciences – Institut de France - Château Observatoire Abbadia - 64700 Hendaye, France.\\
$^{101}$ Brasserie Meteor, 6 Rue Lebocq 67270 Hochfelden, France.\\
$^{102}$ Astro-Centre Yonne, 77 bis rue émile tabarant Laroche 89400 St Cydroine, France.\\
$^{103}$ Communauté de Communes du Canton d'Oust 5 chemin de Trésors, 09140   Seix, France.\\
$^{104}$ Société Astronomique de Touraine Le Ligoret 37130    Tauxigny-Saint Bauld, France.\\
$^{105}$ Observatoire de Dax, Rue Pascal Lafitte 40100  Dax, France.\\
$^{106}$ Mairie, 4 Place de l'Église 36230 Saint-Denis-de-Jouhet, France.\\
$^{107}$ Department of Physics and Astronomy, University of Western Ontario, London, Ontario, N6A 3K7, Canada.\\
$^{108}$ Lycée Xavier marmier- 25300 Pontarlier, France.\\
$^{109}$ Université de Technologie de Troyes (UTT) 10004 Troyes, France.\\
$^{110}$ Lycée Polyvalent d'Etat, 20137        Porto-Vecchio, France \\
$^{111}$ Communauté de communes de Bassin d'Aubenas 07200 Ucel. France.\\
$^{112}$ Service hydrographique et océanographique de la marine (Shom), 29200 Brest, France.\\
$^{113}$ laboratoire Morphodynamique Continentale et Côtière (M2C), UMR6143, Université de Caen, 14000 Caen, France. \\
$^{114}$ FRIPON-Austria.\\
$^{115}$ GEPI, Observatoire de Paris, PSL Research University, CNRS, 61 Avenue de l'Observatoire, 75014 Paris, France.\\
$^{116}$ Pôle d'accueil universitaire Séolane, 04400 Barcelonnette, France.\\
$^{117}$ Observatoire Populaire de Laval - Planétarium 53320 Laval, France.\\
$^{118}$ Muséum national d'Histoire naturelle, 75005 Paris, France.\\
$^{119}$ Institut de radioastronomie millimétrique, Université Grenoble Alpes  38400 Saint-Martin-d'Hères, France.\\
$^{120}$ Laboratoire GSMA, UMR CNRS 7331, Université de Reims Champagne-Ardenne, 51687 Reims, France.\\
$^{121}$ École d'ingénieurs en Sciences Industrielles et Numérique - Université de Reims Champagne-Ardenne 08000 Charleville-Mézières, France.\\
$^{122}$ Lycée Robespierre, 62000 Arras, France.\\
$^{123}$ Cité du Volcan, Bourg Murat 97418 Plaine des Cafres 97421, Ile de La Réunion, France.\\
$^{124}$ Observatoire des Makes, Les Makes, 97421 Saint-Louis, Ile de la La Réunion, France. \\
$^{125}$ Observatoire du Maido, OSU-Réunion, CNRS, 97460 Saint Paul, Ile de la Réunion, France.\\
$^{126}$ FRIPON Vigie-Ciel, Ile de la Réunion, France.\\
$^{127}$ Observatoire du Pic des Fées, Mont des oiseaux 83400 Hyères, France.\\
$^{128}$ Association AstroLab 48190 Le Bleymard, France.\\
$^{129}$ E.P.S.A. Etablissement public des stations d'altitude 64570 La Pierre Saint Martin, France.\\
$^{130}$ Observatoire de Boisricheux 28130 Pierres, France.\\
$^{131}$ Association d'astronomie du pays Royannais: Les Céphéides 17200 Royan, France.\\
$^{132}$ Observatoire de Rouen 76000 Rouen, France.\\
$^{133}$ Communauté de Communes du Pays Châtillonnais 21400 Châtillon-sur-Seine, France.\\
$^{134}$ Space sciences, Technologies Astrophysics Research (STAR) Institute, Université de Liège, Liège B-4000, Belgium.\\
$^{135}$ IUT Chalon sur Saône, 71100 Chalon-sur-Saône, France.\\
$^{136}$ 136 Kepler-Gesellschaft, 71263 Weil der Stadt, Germany\\
$^{137}$ Royal Belgian Institute for Space Aeronomy, Brussels, Belgium.\\
$^{138}$ Lycée Polyvalent Robert Garnier, 72405 La Ferté Bernard - France\\
$^{139}$ Observatoire des Pléiades, Les Perrots, 26760 Beaumont lès Valence, France.\\
$^{140}$ CEA, DAM, DIF, F-91297, Arpajon, France.\\
$^{141}$ Uranoscope, Avenue Carnot 7, 77220 Gretz-Armainvilliers, France.\\
$^{142}$ Observatoire de Haute Provence-Institut Pythéas, CNRS - Aix-Marseille Université, 04870 Saint Michel l'Observatoire, France.\\
$^{143}$ High Enthalpy Flow Diagnostics Group, Institut für Raumfahrtsysteme, Universität Stuttgart, D–70569 Stuttgart, Germany.\\
$^{144}$ Club Ajaccien des Amateurs d'Astronomie, Centre de recherche scientifique Georges Peri 20000 Ajaccio, France.\\
$^{145}$ Laboratoire de Planétologie et Géodynamique, UMR6112, CNRS, Université Nantes, Université Angers, Nantes, France.\\
$^{146}$ Laboratoire d’Océanologie et de Géosciences UMR 8187, 62930 Wimereux, France.\\
$^{147}$ Blois Sologne Astronomie 41250 Fontaines-en-Sologne, France.\\
$^{148}$ Planétarium d’Epinal, 88000 Épinal, France.\\
$^{149}$ Institut UTINAM UMR 6213, CNRS, Université Bourgogne Franche-Comté, OSU THETA, 25010 Besançon, France.\\
$^{150}$ Arbeitskreis Meteore e.V, Germany.\\
$^{151}$ La Ferme des Etoiles, 32380 Mauroux, France.\\
$^{152}$ Bibracte, Centre archéologique, 58370 Glux-en-Glenne\\
$^{153}$ Laboratoire Univers et Particules de Montpellier, Universit\'e de Montpellier, UMR-CNRS 5299, 34095 Montpellier Cedex, France\\
$^{154}$ Laboratoire de Planétologie et Géodynamique, UMR 6112, CNRS - Département de Géosciences, Le Mans Université, Le Mans, France.\\
$^{155}$ Récréa Sciences (CCSTI du Limousin)  23200 Aubusson, France.\\
$^{156}$ Centro de Astronom{\'i}a (CITEVA), Universidad de Antofagasta, 1270300 Antofagasta, Chile.\\
$^{157}$ Astronomical Institute of the Romanian Academy, Bucharest, RO-040557, Romania.\\
$^{158}$ Planetarium Pythagoras Via Margherita Hack, 89125 Reggio Calabria, RC, Italy.\\
$^{159}$ Observatoire astronomique jurassien, Chemin Des Ecoles 21, CH-2824 Vicques, Switzerland.\\
$^{160}$ Le Don Saint 19380 Bonnet Elvert, France.\\
$^{161}$ Mairie, Le Village, 66360 Mantet, France.\\
$^{162}$ Planetarium de Bretagne, 22560 Pleumeur Bodou, France.\\
$^{163}$ Club St Quentin Astronomie, 02100 Saint Quentin, France.\\
$^{164}$ MAYA (Moulins Avermes Yzeure Astronomie) 03000 Moulins, France.\\
$^{165}$ Laboratoire de Géologie de Lyon : Terre, Planète, Environnement, UMR CNRS 5276 (CNRS, ENS, Université Lyon1), Lyon, France.\\
$^{166}$ IRAP, Université de Toulouse, CNRS, UPS, CNES, Toulouse, France.\\
$^{167}$ Institut de Ciències del Cosmos (ICC-UB-IEEC),  1, Barcelona E-08028, Spain.\\
$^{168}$  Parc Astronòmic Montsec - Ferrocarrils de la Generalitat de Catalunya, Ager E-25691, Spain.\\
$^{169}$ Parc naturel régional des Landes de Gascogne, 33380 Belin-Béliet, France.\\
$^{170}$ Université Côte d'Azur, Observatoire de la Côte d'Azur, CNRS, Laboratoire Lagrange,UMR 7293, CNRS, Université de Nice Sophia-Antipolis, Nice, France\\
$^{171}$ Association Pierre de Lune, 87600 Rochechouart, France.\\
$^{172}$ Hotel De Ville, Plaine De Cavarc, 47330 Cavarc, France.\\
$^{173}$ Planète et Minéral Association, 16 rue d’aussières 11200 Bizanet, France.\\
$^{174}$ Marie, 85120 La Chapelle aux Lys, France.\\
$^{175}$ Mairie de Saint-Lupicin, 2 Place de l'Hôtel de ville, Saint-Lupicin, 39170 Coteaux du Lizon\\
$^{176}$ Planétarium et Centre de Culture Scientifique et Technique (le PLUS), 59180 Cappelle la Grande, France.\\
$^{177}$ Université de Bordeaux, CNRS, LOMA, 33405 Talence, France.\\
$^{178}$ Instituto de Astrofísica, PUC, Santiago, Chile.\\
$^{179}$ Club Alpha Centauri, 11240     Cailhavel, France.\\
$^{180}$ Lycée Pierre Forest, 59600 Maubeuge, France.\\
$^{181}$ Club d'Astronomie Jupiter du Roannais, Mairie de Villerest, 7 Rue du Clos 42300 Villerest, France.\\
$^{182}$ Planétarium du Jardin des Sciences, 67000 Strasbourg, France.\\
$^{183}$ Collège Robert Doisneau: association Sirius 57430 Sarralbe, France.\\
$^{184}$ West University of Timisoara, Faculty of Mathematics and Computer Science, Romania.\\
$^{185}$ Romanian Society for Cultural Astronomy, Romania.\\
$^{186}$ Romanian Society for Meteors and Astronomy (SARM), Romania. \\
$^{187}$ La Torre del Sole, Via Caduti sul Lavoro 2, 24030 Brembate di Sopra, BG, Italy.\\
$^{188}$ Associazione Astrofili Bisalta Via Gino Eula 23, 12013 Chiusa di Pesio, CN, Italy.\\
$^{189}$ Department of Earth and Environmental Sciences, The University of Manchester, UK.\\
$^{190}$ DarkSkyLab, 3 rue Romiguières, 31000 Toulouse, France \\
$^{191}$ School of Physical Sciences, The Open University, UK.\\
$^{192}$ European Space Agency, Oxford, UK.\\
$^{193}$ Amgueddfa Cymru - National Museum Wales, Cardiff, Wales.\\
$^{194}$ lycée Gustave Flaubert, La Marsa, Tunisia.\\
$^{195}$ FRIPON - Tunisia.\\
$^{196}$ Observatoire François-Xavier Bagnoud, 3961 St-Luc, Switzerland.\\
$^{197}$ LFB - Lycée français de Barcelone - Bosch i Gimpera 6-10 - 08034 Barcelona, Spain.\\
$^{198}$ Meteoriti Italia APS   Via Fusina 6, 32032 Feltre, BL, Italy.\\
$^{199}$ Associazione Sky Sentinel      Via Giovanni Leone 36, 81020 San Nicola la Strada CE, Italy.\\
$^{200}$ Chair of Astronautics, TU Munich, Germany.\\
$^{201}$ Herrmann-Lietz-Schule, Spiekeroog, Germany.\\
$^{202}$ Förderkreis für Kultur, Geschichte und Natur im Sintfeld e. V., Fürstenberg, Germany.\\
$^{203}$ EUC Syd, Sønderborg, Denmark.\\
$^{204}$ Bundesanstalt für Geowissenschaften und Rohstoffe, Hannover, Germany.\\
$^{205}$ Deutschen Schule Sonderburg, Denmark.\\
$^{206}$ Observatoire d'Alger, CRAAG, Route de l'Observatoire, Alger, Algéria.\\
$^{207}$ Physical-Geographic and Environmental Quality Monitoring Research Station Mădârjac - Iași, Faculty of Geography and Geology, ”Alexandru Ioan Cuza” University of Iași, RO-700506, Romania.\\
$^{208}$ Planetarium and Astronomical Observatory of the Museum “Vasile Pârvan” Bârlad, RO - 731050, Romania.\\
$^{209}$ Galați Astronomical Observatory of the Natural Sciences Museum Complex, 800340, Galați, Romania.\\
$^{210}$ BITNET Research Centre on Sensurs and Systems,, Cluj-Napoca, RO-400464, Romania.\\
$^{211}$ Romanian Academy Timisoara Branch, Astronomical Observatory Timisoara, 300210 Timisoara, Romania.\\
$^{212}$ San Pedro de Atacama Celestial Explorations, Casilla 21, San Pedro de Atacama, Chile.\\
$^{213}$ Institut de Technologie Nucléaire Appliquée, Laboratoire Atomes Laser, Université Cheikh Anta Diop, Dakar, Senegal.\\
$^{214}$ Centre d'Ecologie et des Sciences de la Conservation (CESCO), MNHN, CNRS, Sorbonne Université, Paris, France.\\
$^{215}$ Mairie de Zicavo, Quartier de l'Église, 20132 Zicavo, France.\\
$^{216}$ Club Pégase, amicale laïque de Saint-Renan, Rue de Kerzouar. 29290 Saint-Renan, France.\\
$^{217}$ Club d'Astronomie de Rhuys, Château d'eau de Kersaux, 56730 Saint-Gildas-de-Rhuys, France.\\
$^{218}$ L2n, CNRS ERL 7004, Université de Technologie de Troyes, 10004 Troyes, France.\\
$^{219}$ Mairie, 12, rue des Coquelicots 12850 Onet-le-Château, France.\\
$^{220}$ Planetarium and Astronomical Observatory of the Museum “Vasile Pârvan” Bârlad, Romania.\\
$^{221}$ Romanian Academy, Astronomical Institute, Astronomical Observatory Cluj,  Cluj-Napoca, Romania.\\
$^{222}$ Secci\'on F\'isica, Departamento de Ciencias, Pontificia Universidad     Cat\'olica del Per\'u, Apartado 1761, Lima, Per\'u\\
$^{223}$ Direction du Patrimoine et des musées Conseil départemental de la Manche - 50050 Saint-Lô, France.\\
$^{224}$ UPJV, Universit\'e de Picardie Jules Verne, 80080 Amiens, France.\\
$^{225}$ IPGS‐EOST, CNRS/University of Strasbourg,  Strasbourg, France.\\
$^{226}$ Mairie de Cailhavel, 11240 Cailhavel, France\\
$^{227}$ Club Alpha Centauri, MJC, 11000 Carcassonne, France.\\
$^{228}$ Universidad Católica del Norte, 0610, Antofagasta, Chile.\\
$^{229}$ Millennium Institute for Astrophysics MAS, Av. Vicuña Mackenna 4860, Santiago, Chile.\\
$^{230}$ American Association of Variable Stars Observers, USA\\
$^{231}$ Institute of Space Sciences (CSIC), Campus UAB, Facultat de Ciències,  08193 Bellaterra, Barcelona, Catalonia, Spain.\\
$^{232}$ Institut d’Estudis Espacials de Catalunya (IEEC), 08034 Barcelona, Catalonia, Spain.\\
$^{233}$ Comisi\'on Nacional de Investigaci\'on y Desarrollo Aeroespacial del Per\'u, CONID9, San Isidro Lima, Per\'u.\\
$^{234}$ Università di Firenze - Osservatorio Polifunzionale del Chianti       Strada Provinciale Castellina in Chianti, 50021 Barberino Val D'elsa, FI, Italy.\\
$^{235}$ Associazione Astrofili Urania  Località Bric del Colletto 1, 10062 Luserna San Giovanni, TO, Italy.\\
$^{236}$ Associazione Culturale Googol  Via Filippo Brunelleschi 21, 43100 Parma, PR, Italy.\\
$^{237}$ Osservatorio Astronomico di Capodimonte        Salita Moiariello 16, 80131 Napoli, NA, Italy.\\
$^{238}$ Fondazione GAL Hassin - Centro Internazionale per le Scienze Astronomiche, 90010 Isnello, Palermo, PA, Italy.\\
$^{239}$ Università del Salento - Dipartimento di Matematica e Fisica  Via Per Arnesano, 73100 Lecce, LE, Italy.\\
$^{240}$ Gruppo Astrofili Monti Lepini - Osservatorio Astronomico e Planetario di Gorga        00030 Gorga, RM, Italy.\\
$^{241}$ Associazione Astrofili di Piombino - Osservatorio Astronomico Punta Falcone Punta Falcone, Località Falcone, 57025 Piombino, LI, Italy.\\
$^{242}$ CIRA - Centro Italiano Ricerche Aerospaziali   Via Maiorise snc, 81043 Capua, CE, Italy.\\
$^{243}$ Astrobioparco Oasi di Felizzano        Strada Fubine 79, 15023 Felizzano, AL, Italy.\\
$^{244}$ Associazione Astrofili Tethys - Planetario e Osservatorio Astronomico Cà del Monte   Località Ca del Monte, 27050 Cecima, PV, Italy.\\
$^{245}$ GAMP - Osservatorio Astronomico Montagna Pistoise      51028 San Marcello Piteglio, PT, Italy.\\
$^{246}$ Gruppo Astrofili Antares       Via Garibaldi 12, 48033 Cotignola, RA, Italy.\\
$^{247}$ SpaceDys       Via Mario Giuntini 63, 56023 Navacchio di Cascina, PI, Italy.\\
$^{248}$ Associazione Astrofili Tethys - Planetario e Osservatorio Astronomico Cà del Monte   Località Ca' del Monte, 27050 Cecima, PV, Italy.\\
$^{249}$ Liceo Statale "Arturo Issel"   Via Fiume 42, 17024 Finale Ligure, SV, Italy.\\
$^{250}$ Università di Camerino - Scuola di Scienze e Tecnologie, sezione Geologia        Via Gentile III da Varano, 62032 Camerino, MC, Italia.\\
$^{251}$ Osservatorio Astrofisico R.P.Feynman   73034 Gagliano del Capo, LE, Italy.\\
$^{252}$ Manca  Osservatorio Astronomico di Sormano     Località Colma di, 22030 Sormano, CO, Italy.\\
$^{253}$ Associazione Astronomica del Rubicone  Via Palmiro Togliatti 5, 47039 Savignano sul Rubicone, FC, Italy.\\
$^{254}$ Università degli Studi di Firenze - Dipartimento di Fisica e Astronomia       Via Sansone 1, 50019 Sesto Fiorentino, FI, Italy.\\
$^{255}$ IIS "E. Fermi" di Montesarchio Via Vitulanese, 82016 Montesarchio, BN, Italy.\\
$^{256}$ Liceo Scientifico Statale "G.B. Quadri"        Viale Giosuè Carducci 17, 36100, Vicenza, VI, Italy.\\
$^{257}$ Università degli Studi di Trento - Dipartimento di Ingegneria Civile, Ambientale e Meccanica  Via Mesiano 77, 38123 Trento, TN, Italy.\\
$^{258}$ Osservatorio Astronomico Sirio Piazzale Anelli, 70013 Castellana Grotte, BA, Italy.\\
$^{259}$ Museo del Cielo e della Terra  Vicolo Baciadonne 1, 40017 San Giovanni in Persiceto, BO, Italy.\\
$^{260}$ Gruppo Astrofili Montelupo Fiorentino  Piazza Vittorio Veneto 10, 50056 Montelupo Fiorentino, FI, Italy.\\
$^{261}$ Università del Piemonte Orientale - Dipartimento di Scienze e Innovazione Tecnologica     Viale Teresa Michelin 11, 15121 Alessandria, AL, Italy.\\
$^{262}$ Osservatorio Astronomico Giuseppe Piazzi       Località San Bernardo, 23026 Ponte in Valtellina, SO, Italy.\\
$^{263}$ Liceo Scientifico Statale "P. Paleocapa"       Via Alcide de Gasperi 19, 45100 Rovigo, RO, Italy.\\
$^{264}$ Osservatorio Astronomico Bobhouse      Via Giuseppe Tomasi P.pe di Lampedusa 9, 90147 Palermo, PA, Italy.\\
$^{265}$ Observatoire de la grande vallée, 16250, Etriac, France\\
$^{266}$ South African Astronomical Observatory, University of Cape Town, South Africa.\\
$^{267}$ Departamento de Matemáticas y Computación. Universidad de La Rioja, Spain.\\
$^{268}$ Departamento de Estadística, Informática y Matemáticas and Institute for Advanced Materials and Mathematics, Universidad Pública de Navarra, 31006 Pamplona, Spain.\\
$^{269}$ Laboratoire d'Informatique de Paris (LIP6), Sorbonne Universite, CNRS, Paris, France\\
$^{270}$ Space Science and Technology Centre, School of Earth and Planetary Sciences, Curtin University, Perth, WA 6845, Australia. 
$^{271}$ Department of Physics, University of Western Australia, Crawley 6009, Australia
$^{272}$ Australian Research Council Centre of Excellence, OzGrav\\
$^{273}$ ASPA (Association Sénégalaise pour la Promotion de l'Astronomie), Dakar, Sénégal.
\\
\email{Francois.colas@obspm.fr}  }

\date{Received ...; accepted ...}

 
\abstract
   {...}
   {...}
   {...}
   {..}
   {...}
   
\titlerunning{FRIPON fireball network}
\authorrunning{F. Colas et al. }

\maketitle

\section{Abstract}

\textit{Context:} Until recently, camera networks designed for monitoring fireballs worldwide were not fully automated, implying that in case of a meteorite fall, the recovery campaign was rarely immediate. This was an important limiting factor as the most fragile - hence precious - meteorites must be recovered rapidly to avoid their alteration.  

\textit{Aims:} The Fireball Recovery and InterPlanetary Observation Network (FRIPON) scientific project was designed to overcome this limitation. This network comprises a fully automated camera  and radio network deployed over a significant fraction of western Europe and a small fraction of Canada. As of today, it consists of 150 cameras and 25 European radio receivers and covers an area of about $1.5$ $\times$  $10^6$ km$^2$.

\textit{Methods:} The FRIPON network, fully operational since 2018, has been monitoring meteoroid entries since 2016, thereby allowing the characterization of their dynamical and physical properties. In addition, the level of automation of the network makes it possible to trigger a meteorite recovery campaign only a few hours after it reaches the surface of the Earth. Recovery campaigns are only organized for meteorites with final masses estimated of at least 500 g, which is about one event per year in France. No recovery campaign is organized in the case of smaller final masses on the order of 50 g to 100 g, which happens about three times a year; instead, the information is delivered to the local media so that it can reach the inhabitants living in the vicinity of the fall.

\textit{Results:} Nearly 4,000 meteoroids have been detected so far and characterized by FRIPON. The distribution of their orbits appears to be bimodal, with a cometary population and a main belt population. Sporadic meteors amount to about 55\% of all meteors. A first estimate of the absolute meteoroid flux (mag < -5; meteoroid size $\ge\sim$1 cm) amounts to 1,250/year/$10^6$ km$^2$. This value is compatible with previous estimates. Finally, the first meteorite was recovered in Italy (Cavezzo, January 2020) thanks to the PRISMA network, a component of the FRIPON science project.
   
\section{Introduction}

The study of the physical and dynamical properties of interplanetary matter, such as interplanetary dust particles (IDPs), meteoroids, asteroids, comets, is crucial to our understanding of the formation and evolution of the solar system. This matter exists in many sizes, from micron-sized dust grains to several hundred kilometer-sized bodies. Whereas the largest bodies are routinely studied via Earth-based telescopic observations as well as less frequent interplanetary missions, the smallest bodies (diameter $\leq$10 m) are for the most part only observed and characterized when they enter the Earth's atmosphere as their entry generates enough light to be recorded by even the simplest types of cameras; the smaller particles are called meteors and the larger bodies are fireballs.



\begin{figure*}[!]
\includegraphics[width=1\hsize]{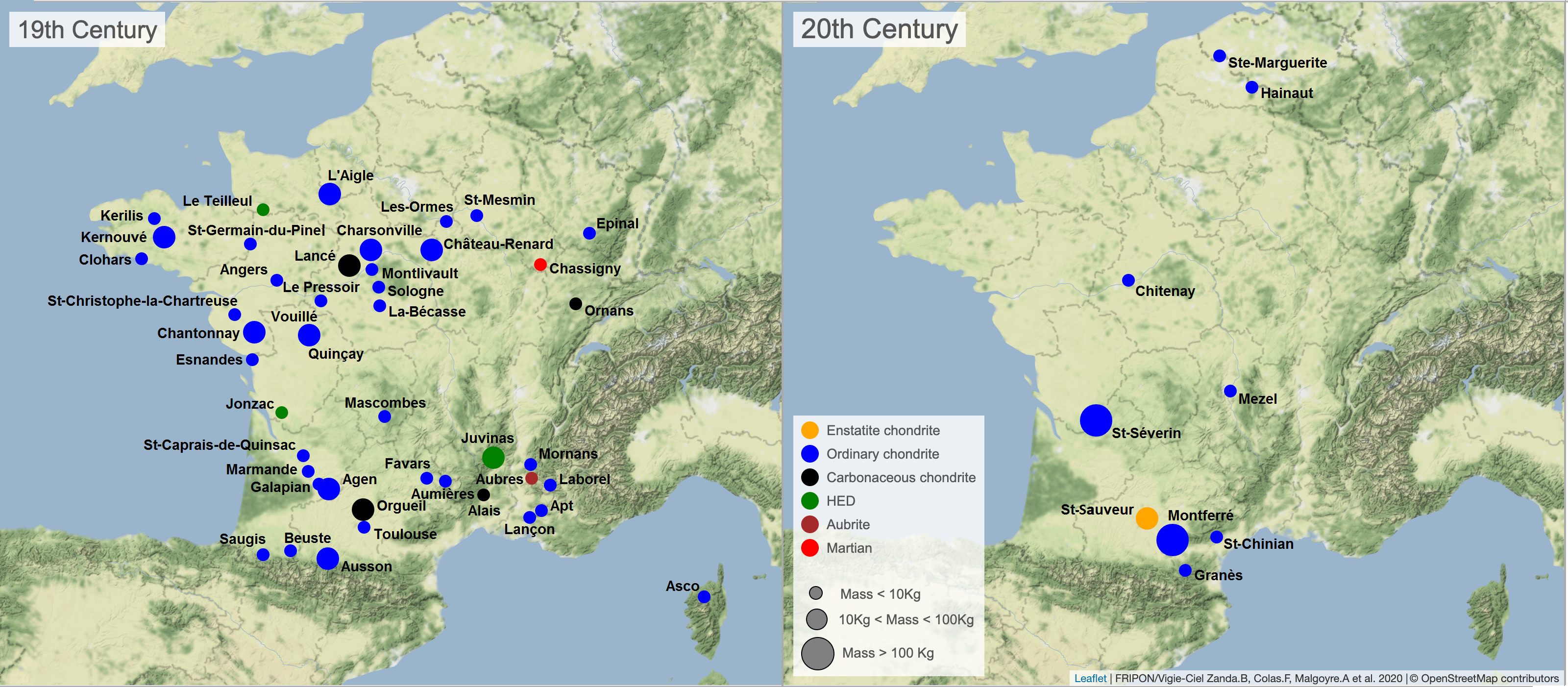}
\caption{In nineteenth century France, 45 meteorites were recovered after their fall was observed, a number that fell by a factor of 5 in the twentieth century.  Even in the nineteenth century, witnessed falls were not randomly distributed. They were mostly located in the great river plains (Seine and Loire in the northwest, Garonne in the southwest, and Rhône valley in the southeast). In these regions, the population was denser, the view is free of obstacles (such as mountains), and the skies are often clear. The striking difference between the two centuries illustrates the need for distributed observers for meteorite recovery. Rural populations have declined because of urbanization in the twentieth century. A camera network such as FRIPON can monitor atmospheric entries and take over that role that was previously played by human observers. However, trained human eyes are still required to recover the meteorites; this is the aim of the Vigie-Ciel citizen science program \citep{2015EPSC...10..604C}.}
\label{Fig:ChutesFrance}
\end{figure*}

We know that $\sim$100 tons of extraterrestrial material collide with the Earth daily, mostly as small particles less than 0.2 mm in size (\citealt{2006mess.book..869Z}, \citealt{2019LPI....50.1968R}). At present, these small particles, called IDPs, are actively being collected in the stratosphere, from polar ices \citep{2007AdSpR..39..605D}, and within impact features on spacecraft \citep{2020JSpRo..57..160M}. For such particles, the stratospheric collections provide the least contaminated and heated samples. At the other end of the size distribution of extraterrestrial material colliding with the Earth, meteorites are fragments that have survived the passage through the atmosphere without internal chemical alteration, which have been recovered at the surface of the Earth. To date, all known meteorites are pieces of either asteroids, the Moon, or Mars, with asteroidal fragments dominating the flux of material, whereas IDPs originate mostly from comets and possibly from asteroids (\citealt{1996M&PS...31..394B, 2015ApJ...806..204V}). The most detailed information on the processes, conditions, timescales, and chronology of the early history of the solar system (e.g., \citealt{2019ApJ...875...30N, 2019GeCoA.262...92K} and references therein), including the nature and evolution of the particles in the pre-planetary solar nebula, has so far come from the study of all these extraterrestrial materials. Recovering intact samples of such materials is therefore a critical goal of planetary studies.

However, we are not very efficient at recovering the meteorites that hit the Earth. Estimates based on previous surveys \citep{1996MNRAS.283..551B} and on collected falls [Meteoritical Bulletin database\footnote{https://www.lpi.usra.edu/meteor/metbull.php}] indicate that, for meteorites with masses greater than 100 g, probably less than 1 in 500 that fall on Earth are currently recovered. In addition, taking France as an example, recovery rates were significantly higher in the nineteenth century than they are now: 45 meteorites were observed to fall and found on the ground in the nineteenth century, whereas they were 5 times fewer in the twentieth century (Fig.~\ref{Fig:ChutesFrance}), showing that there is at present a large potential for improvement. Hot and cold deserts are privileged dense collection areas, but most meteorites are found hundreds and up to millions of years after their fall \citep{2016M&PS...51..468H, 2019Geo....47..673D}. They have thus been exposed to terrestrial alteration, which has partly obliterated the scientific information they contain. Also, the critical information regarding their pre-atmospheric orbit is no longer available.

The most efficient approach for recovering freshly fallen meteorites is to witness their bright atmospheric entry via dense (60-120~km spacing) camera and radio networks. These networks make it possible to accurately calculate their trajectory from which both their pre-atmospheric orbit and their fall location (with an accuracy on the order of a few hundred meters) can be constrained.

Records of incoming meteorites started with the appearance of photographic plates at the end of the twentieth century. A first attempt to observe incoming bolides was made in the United States and consisted of a small camera network that was operated between 1936 and 1951
\citep{10.2307/984939}, but it was  only in the middle of the twentieth century that the first fireball observation networks were developed with the aim of recovering meteorites. Two such networks were established in the 1960s. The first was the Prairie Network \citep{1965SAOSR.173.....M} in the center of the United States, which remained operational from 1964 to 1975. This network comprised 16 stations located 250 km apart. Only one meteorite was recovered thanks to this network (Lost City, 1970; \citealt{1971JGR....76.4090M}). The low efficiency of the Prairie Network, despite the large area it covered ($750,000$  km$^2 $) mainly resulted from the low efficiency of the photographic plates, the large distance between the stations, and the slow pace of the data reduction process. 

The European Fireball Network (EFN) was also developed in the 1960s, under the guidance of the Ondrejov Observatory, following the recovery of the  P{\v r}{\'{\i}}bram meteorite in 1959 \citep{1960BAICz..11..164C}. It is still active, currently covers $1$ $\times$ $10^6$  km$^2$ with about 40
cameras, (\citealt{1998M&PS...33...49O}) and benefits from modern equipment. So far, this network has enabled the recovery of nine meteorites (Table~\ref{tab:METEORITE-ORBITE}).

In 1971, the Meteorite Observation and Recovery Project (MORP) project was established over part of Canada and led to the recovery of the Innisfree meteorite \citep{1978JRASC..72...15H}.
The modern digital camera extension of this network, called the Southern Ontario Meteor Network, led to the recovery of the Grimsby meteorite \citep{2011M&PS...46..339B}. The MORP project comprises 16 cameras and covers a surface area of $700,000$ km$^2$. Other networks using photographic techniques have also been developed, such as the Tajikistan Fireball Network \citep{2015SoSyR..49..275K}, which consists of 5 cameras and covers 11\,000 km$^2$. 
However, none of these other networks have made it possible to recover meteorites so far. 
We note the existence of other networks such as the SPMN network, which facilitated the recovery of the Villalbeto
de la Peña \citep{2006M&PS...41..505T} and Puerto Lápice \citep{2009M&PS...44..159L} meteorites, as well as the Finnish Fireball Network, which facilitated recovering the Annama meteorite \citep{2014pim4.conf..162G,2015MNRAS.449.2119T}.
Last, the Desert Fireball Network \citep{2012AuJES..59..177B} was implemented in Australia in 2007. This network is based on high-resolution digital cameras and has made it possible to recover four meteorites:
 Bunburra Rockhole in 2007 \citep{2012M&PS...47..163S},
 Mason Gully in 2010 \citep{2016M&PS...51..596D}, Murrili  in 2015 \citep{2016LPICo1921.6265B}, and 
Dingle Dell in 2016 \citep{2018M&PS...53.2212D}.
The success of this network results from the efficiency of the cameras and the size of the network as well as an efficient data reduction and analysis  process \citep{2019ApJ...885..115S}. A method to construct a successful fireball network is discussed in \citet{2017ExA....43..237H}.

As of today, there are 38 meteorites with reliable reconstructed orbits, 22 of which were detected by camera networks (see Table \ref{tab:METEORITE-ORBITE}). Among the remaining 16 meteorites, $14$ are the result of random visual observations such as the Chelyabinsk event (data from security cameras were used for orbit computation;  \citealt{2013Natur.503..235B}) and two meteorites were detected as asteroids before their fall (Almahata Sitta and 2018LA). During the same time interval (1959-2020), 397 meteorites were recovered after their falls were witnessed by eye (Meteoritical Bulletin Database).

The main limitation of current networks is their size. Most of these networks consist of a fairly small number of cameras spread over a comparatively small territory. Altogether, they cover only 2\% of the total surface of the Earth \citep{2020arXiv200401069D}. This implies that the number of bright events per year witnessed by these networks is small and that decades would be necessary to yield a significant number ($\geq$100) of samples.

\begin{table}[ht!]
\caption{Thirty-eight known meteorites with reliable orbit reference discovered by
networks (``N''), visual observations (``V'') or telescopic observations (``T'').
Bibliographic references: 
[1] \citealt{1960BAICz..11..164C}; 
[2] \citealt{1971JGR....76.4090M}; 
[3] \citealt{1981Metic..16..153H}; 
[4] \citealt{2014A&A...570A..39S}; 
[5] \citealt{1994Natur.367..624B}; 
[6] \citealt{1996M&PS...31..502B}; 
[7] \citealt{2003M&PS...38..975B}; 
[8] \citealt{2000Sci...290..320B}; 
[9] \citealt{2003Natur.423..151S}; 
[10] \citealt{2004M&PS...39..625S}; 
[11] \citealt{2006M&PS...41..505T}; 
[12] \citealt{2009M&PS...44..211T}; 
[13] \citealt{2012M&PS...47..163S}; 
[14] \citealt{2010DDA....41.0603C}; 
[15] \citealt{2013M&PS...48.1060F}; 
[16] \citealt{2011M&PS...46..339B}; 
[17] \citealt{2010M&PS...45.1392S}; 
[18] \citealt{2010M&PSA..73.5085H}; 
[19] \citealt{2016M&PS...51..596D}; 
[20] \citealt{2013M&PS...48.1757B}; 
[21] \citealt{2015M&PS...50.1244B}; 
[22] \citealt{2012Sci...338.1583J}; 
[23] \citealt{2014M&PS...49.1388J}; 
[24] \citealt{2013Natur.503..235B}; 
[25] \citealt{2020M&PS...55..376S}; 
[26] \citealt{2015MNRAS.449.2119T}; 
[27] \citealt{2019M&PS...54..699J};  
[28] \citealt{2020arXiv200607151S}; 
[29] \citealt{2018M&PS...53.2212D}; 
[30] \citealt{doi:10.1111/maps.13452}; 
[31] \citealt{2017M&PS...52.1683B}; 
[32] \citealt{2017EPSC...11..995G}; 
[33] \citealt{2017P&SS..143..192S}; 
[34] \citealt{2019M&PS...54.2027B}; 
[35] \citealt{2018RNAAS...2...57D}; 
[36] \citealt{2019ChEG...79l5525B}; 
[37] \citealt{2020...Cavezzo}; 
[38] http://www.prisma.inaf.it/index.php/2020/03/03/the-daylight-fireball-of-february-28-2020/, 
[39] \citealt{2020M&PS...55..231M}.
}


\label{tab:METEORITE-ORBITE}

\centering
\begin{tabular}{c c c c c}
\hline
\hline
Year  & Location & Type & Method & Ref \\
\hline 
1959 & P{\v r}{\'{\i}}bram   &  H5 & N & [ 1]\\
1970 & Lost City             &  H5 & N  & [ 2] \\
1977 & Innisfree             &  L5 & V & [ 3] \\
1991 & Bene{\v{s}}ov         & LL3.5 & N & [ 4] \\
1992 & Peekskill             &  H6 & V & [ 5] \\
1994 & St-Robert          &  H5 & V & [ 6] \\
2000 & Mor{\'a}vka           & H5 & N & [ 7] \\
2000 & Tagish Lake           & C2-ung & V & [ 8] \\
2002 & Neuschwanstein        & EL6 & N & [ 9] \\
2003 & Park Forest           &  L5 & V & [10] \\
2004 & Villalbeto de la Pe{\~n}a & L6 & N & [11] \\
2007 & Cali                  & H/L4 & V & [12] \\
2007 & Bunburra Rockhole     & Eucrite & N & [13] \\
2008 & Almahata Sitta        & Ureilite & T & [14] \\
2008 & Buzzard Coulee        &  H4 & V & [15] \\
2009 & Grimsby               &  H5 & N & [16] \\
2009 & Jesenice              &  L6 & N & [17] \\
2009 & Maribo                & CM2 & V & [18] \\
2010 & Mason Gully           &  H5 & N & [19] \\
2010 & Ko{\v{s}}ice          &  H5 & N & [20] \\
2011 & Kri{\v{z}}evci        &  H6 & N & [21] \\
2012 & Sutter's Mill         &   C & V & [22] \\
2012 & Novato                &  L6 & N & [23] \\
2013 & Chelyabinsk           & LL5 & V & [24] \\
2014 & {\v{Z}}{\v{d}}{\'a}r nad S{\'a}zavou & LL5 & N & [25] \\
2014 & Annama                &  H5 & N & [26] \\
2015 & Creston               &  L6 & N & [27] \\
2015 & Murrili               &  H5 & N & [28] \\
2016 & Dingle  Dell          & LL6 & N & [29] \\
2016 & Dishchii'bikoh        & LL7 & V & [30] \\
2016 & Stubenberg            & LL6 & N & [31] \\
2016 & Osceola               &  L6 & V & [32] \\
2016 & Ejby                  & H5/6 & N & [33] \\
2018 & Hamburg               & H4 & V & [34] \\
2018 & 2018 LA               & --- & T & [35] \\
2019 & Renchen               & L5-6 & N & [36] \\
2020 & Cavezzo               & --- & N & [37]\\
2020 & Novo Mesto            & L6 & V & [38] \\
2020 & Ozerki                & L6 & V & [39] \\
\hline
\\
\end{tabular}
\end{table}

\begin{figure*}[h!]
\label{Fig:NETWORK}
\begin{center}
\includegraphics[width=0.9\hsize]{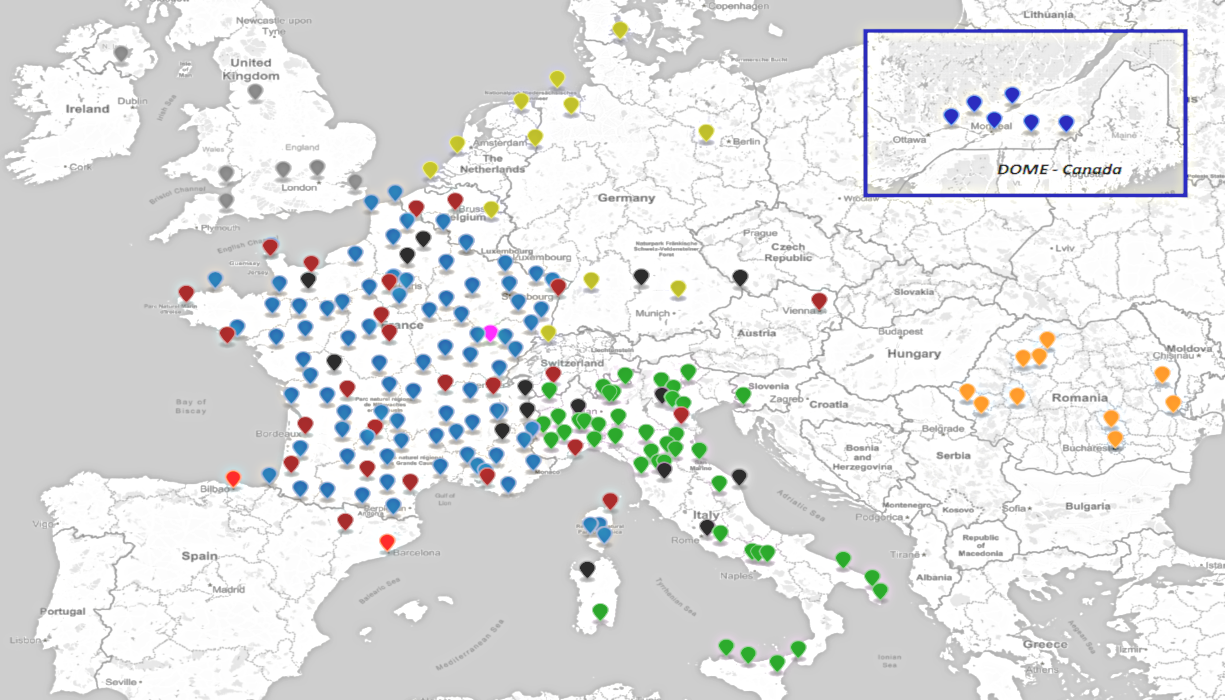}
\end{center}
\caption{FRIPON network map as of end 2019. The color code is the following: \protect\\
1. Blue: FRIPON-France, optical stations.
2. Red: Coupled optical camera and radio receiver stations.
3. Black: Stations under development.
4. Green: PRISMA (Italy).
5. Light Orange: MOROI (Romania).
6. Yellow: FRIPON-Belgium/Neterlands/Germany/Denmark.
7. Gray: SCAMP (United Kingdom).
8. Dark blue: DOME (Canada).
9. Dark Orange: SPMN (Spain).
10. Pink: GRAVES radar.} 
\end{figure*}

The Fireball Recovery and InterPlanetary
Observation Network  (FRIPON)\ scientific project was designed to contribute to this global effort to recover fresh meteorites. It comprises a network deployed over a large fraction of western Europe and a small fraction of Canada (see Fig.~\ref{Fig:NETWORK}). As of today, this network consists of 150 cameras and 25 receivers for radio detection and covers an area of $1.5$ $\times$  $10^6$ km$^2$ (section~\ref{sec:NETWORK_SIZE}). The FRIPON network is coupled in France with the Vigie-Ciel citizen science program, the aim of which is to involve the general public in the search for meteorites in order to improve their recovery rate. In the present paper, we first describe the technology of the FRIPON network and its architecture, and finally we give the first results obtained after four years of observations and report on the first meteorite recovery in Italy\footnote{Discovered from observations by the PRISMA network, a component of the FRIPON network.} \citep{2020...Cavezzo}.

\section{ FRIPON Science Project }

\label{sec:NETWORK_SIZE}

\subsection{General description of the network}

The FRIPON science project was originally designed by a core team of six French scientists from the Paris Observatory (IMCCE), the French National Museum of Natural History (MNHN-IMPMC), Universit\'e Paris-Saclay (GEOPS), and Aix-Marseille University (LAM / CEREGE / OSU Pyth\'eas) to: i) monitor the atmospheric entry of fireballs, that is, interplanetary matter with typical sizes greater than $\sim$1 centimeter; ii) characterize their orbital properties to constrain both their origin and fall location; and iii) recover freshly fallen meteorites. 
This project benefited from a grant from the French National research agency (Agence Nationale de la Recherche: ANR) in 2013 to install a network of charged coupled device (CCD) cameras and radio receivers to cover the entire French territory. Specifically, the grant was used to design the hardware (section~\ref{sec:section-hardware}), building on experience gained from previous networks; develop an efficient and automatic detection and data reduction pipeline (section~\ref{sec:reduction-pipeline}); and build centralized network and data storage architectures (section~\ref{sec:data-storage}). The FRIPON project is designed as a real-time network with the aim of triggering a field search within the 24 h that follow the fall in order to recover fresh meteorites. As of today, FRIPON-France consists of 105 optical all-sky cameras and 25 receivers for radio detection. These assets are homogeneously distributed over the territory, although the radio network is slightly denser in the south of France (Fig.~\ref{Fig:NETWORK}).

Starting from 2016, scientists from neighboring countries were interested in joining the scientific project through the use of the FRIPON-France\footnote{FRIPON-France is also known as  FRIPON-Vigie-Ciel, in order to bring to the fore its citizen science component in France.} hardware, software, and infrastructure. This was the case for Italy (PRISMA network; \citealt{2016pimo.conf...76G, 2019A&A...626A.105B}), Germany (FRIPON-Germany), Romania (FRIPON-MOROI network; \citealt{2019RoAJ...29..189A, 2018RoAJ...28...57N}), the United Kingdom (FRIPON-SCAMP), Canada (FRIPON-DOME), the Netherlands (FRIPON-Netherlands), Spain (FRIPON-Spain), Belgium (FRIPON-Belgium), and Switzerland (FRIPON-Switzerland). Single FRIPON cameras were also made available to the following countries to initiate new collaborations: Austria, Brazil, Chile, Denmark, Mexico, Morocco, Peru, and Tunisia.
As of today, 150 cameras, using FRIPON technology, and 25 radio receivers are operational around the world (see Fig.~\ref{Fig:NETWORK}). 

The FRIPON science project regroups all the above-mentioned national networks, with all the cameras monitored and remotely controlled by the Service Informatique  Pyth\'eas (SIP; Aix-Marseille University, France), which maintains the whole network with the support of the scientific team. All the data from the FRIPON network are stored and processed in Marseille. The data processing consists of monthly astrometric and photometric reduction of the calibration images and daily processing of multi-detections. 
Two databases host the data. One stores the raw data and the other stores higher-level, processed data, such as orbits and trajectories.
These data are available to all coinvestigators of the network\footnote{https://fireball.fripon.org}.
On request, national data can be sent to a different reduction pipeline for alternate processing and storage\footnote{For example, PRISMA data are also stored at the INAF IA2 (Italian Center for Astronomical Archives)   facilities in Trieste \citep{2014ASPC..485..131K} and processed by an independent pipeline (\citealt{2019A&A...626A.105B}, \citealt{2020EPJP..135..255C}).}.

\subsection{Hardware and observing strategy}  
\label{sec:section-hardware}

\subsubsection{Optical cameras} 

Since the early 2000s, digital cameras have been used by all networks that are deployed to monitor fireballs. Two alternate technical solutions are adopted. The first is based on a low-resolution detector (e.g., Southern Ontario Meteor Network; \citealt{2011M&PS...46..339B}), while the second relies on a high-resolution detector (e.g., Desert Fireball Network; \citealt{2012AuJES..59..177B}). The measurements acquired by low-resolution cameras can be accurate enough to compute orbits and strewn fields as long as the network is dense, with numerous cameras. For example, the Southern Ontario Meteor Network, which has been operating in  Canada since 2004, led to the recovery of the Grimsby meteorite \citep{2011M&PS...46..339B}. In the case of the FRIPON network, we followed the philosophy of the Canadian Fireball Network  \citep{2011M&PS...46..339B} as detailed hereafter.

\begin{figure}
\begin{center}
\includegraphics[width=1.0\hsize]{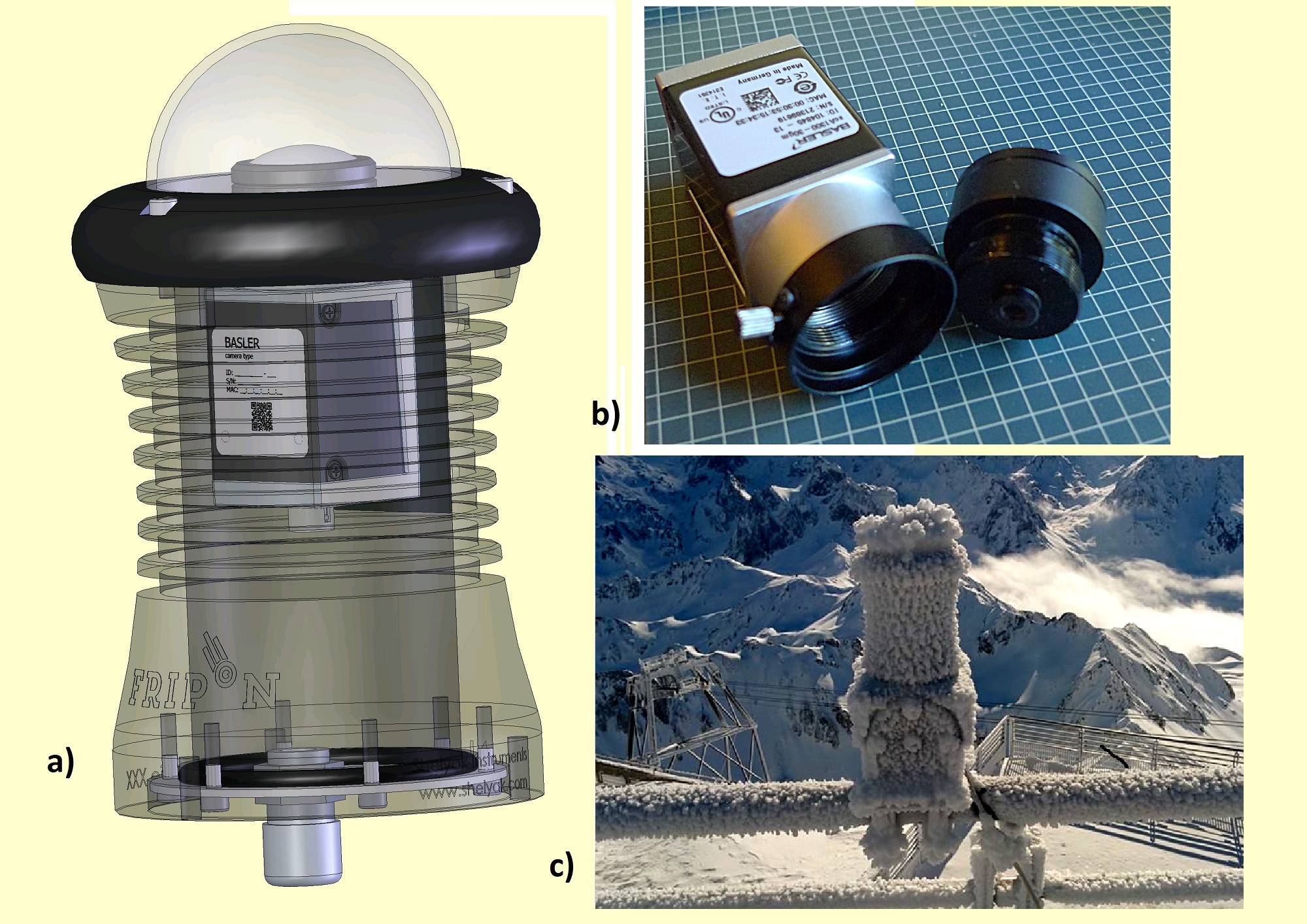}
\end{center}
\caption{Mosaic of technology developed for the FRIPON network: a) Final design of optical detectors$^2$. b) Core device comprising a GigaBit Ethernet camera and fish-eye optics.
 c) FRIPON optical camera installed on the platform of Pic du Midi Observatory (2,876 meters altitude), in use during harsh weather conditions.}
\label{camera}
\end{figure}

We used a CCD Sony ICX445 chip with 1296x964 pixels and a pixel size of 3.75 x  3.75 microns. For the optical design, we used a 1.25 mm focal length F/2 fish-eye camera lens, which leads to a pixel scale of 10 arcmin. Given that fireballs are typically observed at an altitude between 100 km and 40 km, we designed a network with a median distance of 80 km between cameras to perform an optimal triangulation. 
\cite{2019A&A...627A..78J} showed that the astrometric accuracy is on the order of 1 arcmin, equivalent to 30 m at a distance of 100 km. In section~\ref{sec:First-result}, we show that the final accuracy on the trajectory is on the order of 20 m for the position and of 100 m/s for the velocity; this value is required for the identification of meteorite source regions in the solar system as shown by \cite{2018Icar..311..271G}.

The optical device and the CCD were embedded into a special case (Fig. 3) sealed with a transparent dome, thereby allowing us to record full-sky images. Moreover, these cases are equipped with a passive radiator, which serves to release the heat produced by the electronics during the warm periods of the year to minimize CCD dark current.

Each camera is controlled by an Intel NUCi3 computer on which the data are temporarily stored. A single power over ethernet (PoE) cable is used for data transfer and for powering and remotely managing the camera through a TPLINK (TL-SG22110P or 1500G-10PS) switch. Such a solution makes it easy to install the optical station and operate it remotely and to use cables up to 100 meters long between the camera and the computer. Fig.~\ref{camera} shows the design\footnote{Shelyak Instruments, www.shelyak.com} of the camera as well as its installation at the Pic du Midi Observatory.

\subsubsection{Radio receivers}  

In addition to optical observations, we used the powerful signal of the GRAVES radar of the French Air Force. This radar is particularly well adapted for the detection, identification, and tracking of space targets including incoming meteoroids \citep{2005ESASP.587...61M}. Located near Dijon (Burgundy, central eastern France), its four main beams transmit nominally on a half-volume located south of a line between Austria and western France. However, the secondary radiation lobes of the radar make it possible to also detect meteors that disintegrate in the northern part of France. For such observations we do not need as tight a mesh as we do for  the optical network. We have 25 stations with an average distance of 200 km, mainly in France, but also in Belgium, United Kingdom, Italy, Switzerland, Spain, and Austria.  The
GRAVES radar system transmits on 143.050 MHz in a continuous wave (CW) mode 24 hours a day.
A meteoroid entering the E and D layers of the Earth ionosphere produces ions and free electrons generated by the ionization of air and of meteoroid molecules. The free electrons have the property of scattering radio waves according to "back or forward meteor scatter" modes when they are illuminated by a radio transmitter. The FRIPON radio setup is presented in the Appendix.

\subsubsection{Data storage and access}
\label{sec:data-storage}

The FRIPON stations are composed of a Linux minicomputer, a wide-angle camera, and a manageable switch guaranteeing the isolation of the network of the host institute. The installation is done with an automated deployment system based on a USB key.

When connecting to the host, the station establishes a secure VPN tunnel to the central server of the FRIPON project hosted by the information technology department of the OSU Institut Pyth\'eas (SIP) for all cameras and partner networks worldwide. The minicomputer is used for the acquisition and temporary storage of long exposure captures, and detections through the FreeTure open source software \citep{2014pim4.conf...39A} and a set of scripts. 
The data, which include astrometric long exposures images, single detection (stacked images), and multiple detections (both optical and radio raw data) are subsequently transferred to the central server.

The data collected on the server are then indexed in a database. During this operation, visuals are generated. When an optical event groups at least two stations, the FRIPON pipeline is executed to generate the dynamical and physical properties of the incoming meteoroid such as its orbit, its mass and its impact zone.

All the data are made available through a web interface that is accessible to the worldwide community in real time\footnote{https://fireball.fripon.org}. This interface makes it possible to display and download data in the form of an archive that complies with the data policy of the project by means of access right management.

\subsubsection{Detection strategy}

The acquisition and detection software FreeTure  was specifically developed by the FRIPON team and runs permanently on the minicomputers (see \citealt{2014pim4.conf...39A} for a full description). 
The images corresponding to single detections by FreeTure are stored locally and 
a warning (time and location) is sent to the central server in Marseille. 
If at least one other station detects an event within +/- 3 seconds, it is then treated as a "multiple detection". We note that we implemented a distance criterion of less than 190 km to avoid false detections.  This value was determined empirically by manually checking one year of double detections.
This strategy works well during the night, but leads to 30\% of false detections mainly during twilight.

Radio data corresponding to the last week of acquisition are only stored locally.
Only radio data acquired at the time of an optical multi-detection are uploaded from the radio stations to the Marseille data center for processing.

\subsection{Data processing}
\label{sec:reduction-pipeline}

\subsubsection{Optical data}

Scientific optical data are  CCD observations recorded at a rate of 30 frames per second (fps). This acquisition rate is necessary to avoid excessive elongation of the meteor in the images in the case of high speed fireballs.  For example, a typical bolide with an average speed of 40 km/s at 100 km altitude at the zenith leads to  a $20^{\circ}/s$ apparent speed on the sky and to a four pixel elongated trail on the CCD. It is larger than the average width of the point spread function  (PSF; typically 1.8 pixels), but still easy to process for centroid determination. No dark and flatfield corrections are made.

However, almost no reference star is measurable on a single frame with such an acquisition speed, as the limiting magnitude is about zero. It is thus necessary to record images with a longer exposure time for calibration.
We therefore recorded five second exposure images every ten minutes; the goal is to~have a decent signal-to-noise ratio\ (S/N) up to a magnitude of 4.5 and~to only marginally affect detection efficiency. Such a calibration strategy allows the detection of a few thousand calibration stars for a given camera on a clear night. To mitigate the effect of cloudy nights and breakdowns, we computed an astrometric calibration once per month for each station. This works for most cameras as their mounts are rigid. However, we occasionally detected flexible mounts based on the repeated calibrations, which led us to shorten the masts of such stations. 

Calibration procedure uses the ICRF2\footnote{http://hpiers.obspm.fr/icrs-pc/newwww/icrf/index.php} reference frame. The distortion function of the optical system is computed in the topocentric horizontal reference system. This allows for an astrometric solution for stars above 10 degrees of elevation  with an accuracy of 1 arcmin. Our procedure leads to the calculation of the azimuth and the elevation of the bolides in the J2000 reference frame. More details regarding our astrometric calibration procedure can be found in \cite{2019A&A...627A..78J}.

\begin{center}
\begin{figure}[!htbp]
\centering
\includegraphics[width=0.99\hsize]{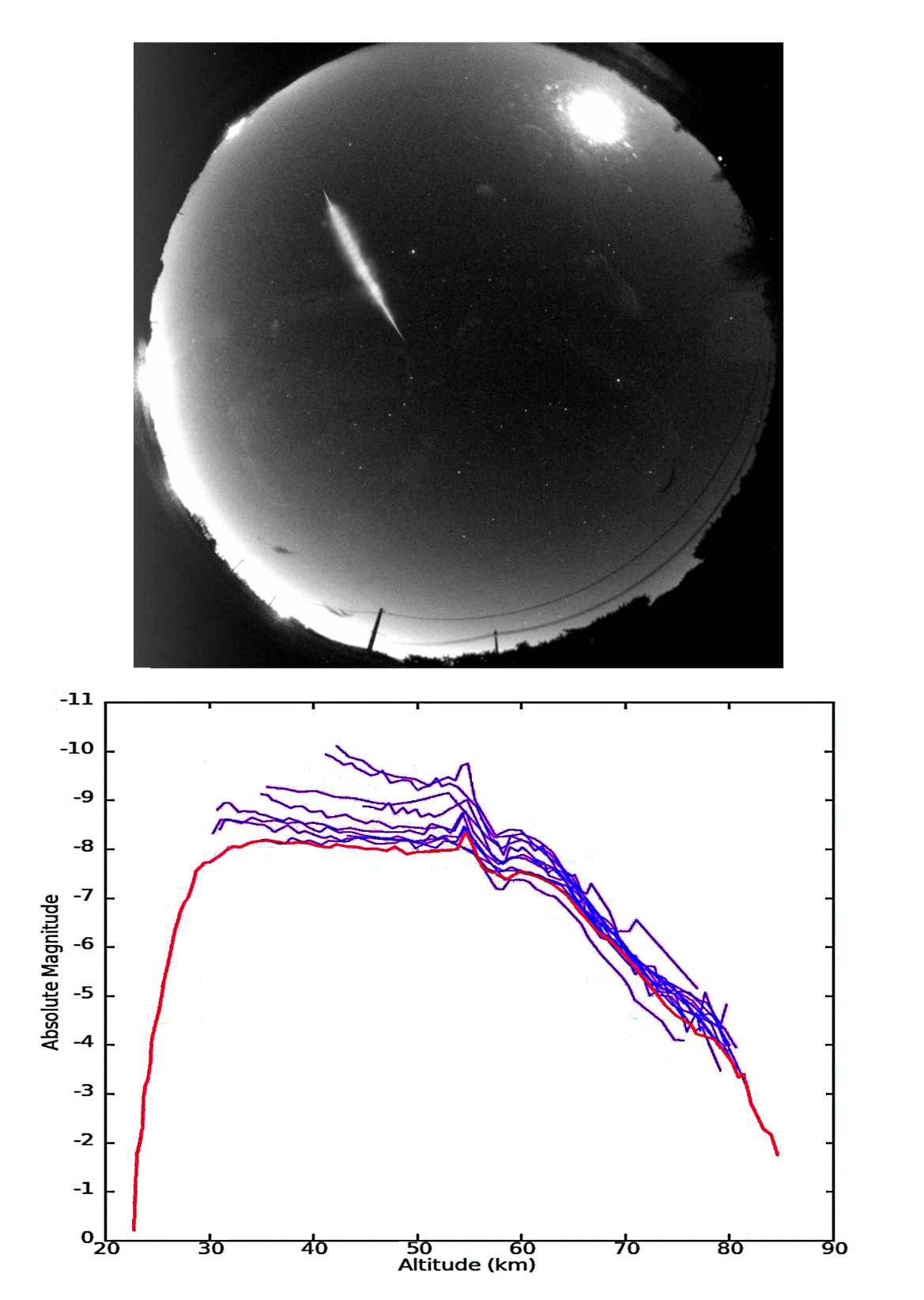}
\caption{ Top: Event on 27 February 2019 seen by the Beaumont-l\`es-Valence FRIPON camera;\protect\\Bottom: Absolute magnitude light curves of the event as seen by 15 cameras; the red curve is Beaumont-l\`es-Valence. It is clear that the saturation limit is around magnitude --8 (all the other light curves fall above this limit). Cameras located further away may be able to measure more non-saturated data, but all the cameras become heavily saturated as the bolide reaches its maximum luminosity.}
\label{Fig:Beaumont-Photometry}
\end{figure}
\end{center}

For the photometric reduction, we used the same frames as for the astrometric calibration, namely the long exposure frames. We then established a correspondence between the observed stars and those present in the Hipparcos catalog \citep{2000PASP..112..961B}.
The following steps are subsequently applied to calculate the absolute magnitude light curve of a meteor, namely: 
i) determination of the flux of an equivalent magnitude 0 star at zenith and the linear extinction function of the air mass for one-month cumulative observation; 
ii) measurement of the bolide flux on individual frames and conversion in magnitude; and
iii) conversion of the meteor magnitude $Mag$ into an absolute magnitude $AMag$, defined as its magnitude at a distance of $100$ km,

\begin{equation}\label{eq:absolu}
AMag_{fireball} = Mag_{fireball}-5 \cdot \log_{10}\left(\frac{d}{100\mathrm{km}}\right)
.\end{equation}

Fig.~\ref{Fig:Beaumont-Photometry} shows the final absolute magnitude light curve of an event recorded by 15 stations on 27 February 2019. We notice that the closest station saturates faster with a -8 magnitude plateau compared to the other cameras. These light curves are saturated at different times, depending on their distance to the bright flight.
For the brightest part of the light curve, a saturation model will be applied in the future. 
At this  point, we point out several limitations of our data reduction procedure as follows:
\begin{itemize}
\item Clouds may partly cover the night sky, which may bias the measure of instrumental magnitudes. 
\item Meteors are mainly detected at small elevations (typically below $30^{\circ}$). These records are therefore affected by nonlinearities of the atmospheric
extinction. 
\item A uniform cloud layer can be the source of an under estimation of bolide magnitude.
\end{itemize}

The first photometric measurements of the FRIPON network are reflected in the histogram of all detections in subsection~\ref{subsec:Quantifying}.
Routines to merge all light curves into one are now under development. As  our data reduction is based on dynamics, the photometric curves are only used at present to detect major events.  

 To summarize, the astrometric reduction allows us to obtain an accuracy of one-tenth pixel or 1 arcmin for meteor measurements. Photometry is at that time only usable for events with an absolute magnitude lower than $-8$ with an accuracy of $0.5$ magnitude.

\subsubsection{Trajectory determination}
\label{sec:trajectory-determination}

Most of our method is described in \cite{2019A&A...627A..78J} and in \cite{2020..sjeanne..thesis} and is only be recalled briefly in this section. Owing to the limited accuracy of the Network Time Protocol (NTP; \citealt{2015PASA...32...14B}), which is  typically 20 ms, we first use a purely geometrical model (without taking into account time) by assuming that the trajectory follows a straight line, after the approach of \cite{1987BAICz..38..222C}. This method allows us to separate the space and time components of our measurements and to overcome the problem of temporal accuracy. We give special attention to global error estimation, which becomes accessible thanks to the large number of cameras involved in most of FRIPON's detections. By comparison, the detections of other networks usually involve fewer cameras, making external biases nonmeasurable and hard to evaluate.  

The density of the FRIPON network makes it possible to observe an event with many cameras (15 in the case of the 27 February 2019 event; see Fig.~\ref{Fig:Beaumont-Photometry}). It is then possible to consider the external astrometric bias of each camera as a random error and to estimate it by a statistical method. Therefore, we developed a modified least-squares regression to fit the data taking into account the internal and external or systematic error on each camera. 

We first estimate the internal error of each camera by fitting a plane passing through the observation station and all the measured points. The average internal error of the cameras amounts to $0.75$ arcmin, which corresponds to 0.07 pixel. 
We also compute a first estimation of the external error by averaging distances between the observed positions of stars and those calculated from the Hipparcos catalog \citep{2000PASP..112..961B} in a neighborhood of $100$ pixels around the meteor. We then compute a global solution using the modified least-squares estimator of the trajectory $\widehat{\mathcal{T}_{\chi^2}}$  given by the minimization of the following sum:

\begin{align}\label{eq:modified_sum_residues}
& &S(\mathcal{T})&= \sum_{i=1}^{n_{cam}} \sum_{j=1}^{n_i} \frac{\epsilon_{ij}(\mathcal{T})^2}{\sigma_i^2 + n_i s_i^2}, 
\end{align}

where $\epsilon_{ij}(\mathcal{T})$ is the residual between the $j^{th}$ measure taken by the $i^{th}$ camera and the trajectory
$\mathcal{T}$, $\sigma_i$ is the internal error of the $i^{th}$ camera, $s_i$ is the systematic error of the $i^{th}$ camera, and $n_i$ is the number of images taken by the $i^{th}$ camera.\\

This method allows us to characterize the systematic errors of our cameras (e.g., a misaligned lens), but not errors such as the location of the camera. To tackle these errors, we compute a first estimate of the trajectory and we compare the residuals with the expected random and systematic errors. If they are larger than expected for a specific camera, we iteratively decrease its weight during the calculation of the trajectory.
The final systematic error is usually on the order of $0.3$ arcmin, which ends the iterative process.

Two geometric configurations lead to important errors or degeneracies in the trajectory determination:  stations located too far from the fireball and stations aligned with the trajectory of the fireball. However, most of the time, the final bright flight straight line trajectory is known with a precision of a few tens of meters. In a second step, all individual data points with time stamps are projected on the straight line to be used afterward for dynamical purposes.

\subsubsection{Orbit, drag, and ablation model}
\label{sec:drag-ablation}

To compute the orbit of the bolide parent body, we need to measure its velocity before it has experienced significant interaction with the upper atmosphere.
This interaction starts well before the bright flight. Therefore, we need a deceleration model to estimate the infinite velocity, even if the deceleration is not measurable, which happens to be the case for many events (especially the high speed events). This problem is complex because physical parameters evolve during atmospheric entry and moreover several parameters are unknown such as drag coefficient, object size, shape, density and strength. Like other teams (\citealt{2016P&SS..120...35L}, \citealt{2014P&SS..103..238B}, \citealt{2019Icar..321..388S}, etc...) we use a simple physical model to fit the bright flight data. 

We used a dynamic model from \cite{1983pmp..book.....B}, equations (\ref{eq:acc}) and (\ref{eq:abl}). This model describes the deceleration and ablation of a meteoroid in an atmosphere based on the following three equations :

\begin{align}\label{eq:acc}
& &\frac{\mathrm{d} V}{\mathrm{d} t} &= -\frac{1}{2} \rho_{atm} V^2 c_d\frac{ S_e}{M_e}\frac{s}{m} \\ \label{eq:abl}
& &\frac{\mathrm{d} m}{\mathrm{d} t} &= -\frac{1}{2}  \rho_{atm} V^3 c_h\frac{ S_e}{H M_e} s \\
& &s&=m^\mu,
\end{align}
 where $c_d$ is the drag coefficient, $c_h$ the heat-transfer coefficient, $H$ is the enthalpy of destruction, $\rho_{atm}$ is the gas density, $m$ is the normalized meteoroid mass, $M_e$ is the pre-entry mass, $s$ is the normalized cross-section area, $S_e$ is the pre-entry cross-section area, $\mu$ is the so-called shape change coefficient.
 The atmospheric gas density $\rho_{atm}$ is taken from the empirical model \emph{NRLMSISE-00} \citep{2016P&SS..120...35L}.
 
These three equations can be rewritten into two independent equations  \citep{2014JTAM...44d..15T}. The equation of motion is written as

\begin{align}
& &\frac{\mathrm{d} V}{\mathrm{d} t} &= -\frac{1}{2} A \rho_{atm} V^2 \mathrm{exp} \left( \frac{B}{A}\left(\frac{V_e^2}{2}-\frac{V^2}{2}\right)\right)
\end{align}

and the equation of mass is written as

\begin{align}
& &m &= \mathrm{exp}\left( \frac{B}{A(1-\mu)}\left(\frac{V^2}{2}-\frac{V_e^2}{2} \right)\right),
\end{align}

where $A$ is a deceleration parameter (in square meters per kilogram) and $B$ is an ablation parameter (in square meters per joule) as follows:

\begin{align*}
& &A &= \frac{c_d S_e}{M_e} & B &= (1-\mu)\frac{c_h S_e}{H M_e} .
\end{align*}

We used our model to fit the positions of each observation that is projected on the trajectory line \citep{2019A&A...627A..78J}.
With this model, the observation of a meteor motion makes it possible to estimate the value of the three parameters $V_e$, $A$, and $B$. Using $A$ and $B$ rather than their ratio $A/B$, which is proportional to the enthalpy of destruction $H$ of the meteoroid (\citealt{2014JTAM...44d..15T}), allowed us
to avoid the numerical singularity when $B$ gets close to zero. \cite{2020..sjeanne..thesis} demonstrated that the least-squares estimators of these three parameters have always defined variances and meaningful values, even in the case of faint meteors. Finally, we computed confidence intervals in the three-dimensional parameter space  ($V_e$, $A$, $B$).\\

\subsubsection{Dark flight}

At the end of the bright flight, a meteoroid is subject only to aerodynamic drag (including winds) and gravity. At this stage, the meteoroid speed is too low to cause ablation (hence dark flight). 

The equation of motion during dark flight is as follows:
\begin{align}
& &\frac{\mathrm{d}\overrightarrow{V}}{\mathrm{d}t}&=\frac{1}{2}A_f(V_w)\rho_{atm}V_{w}^2 \overrightarrow{u_{w}}+\overrightarrow{g}
\end{align}
where $A_f(V_w)$ is the deceleration parameter of the fragment, which depends on the wind velocity (relative to the fragment) $V_w$. 
We used a local atmospheric model of wind retrieved from meteorological offices.

The end of the bright flight simulation gives us the initial conditions of the dark flight motion, namely the initial position, speed, and acceleration of the fragment. 
The initial condition of acceleration gives us a definition of $A_{f_0}$, the limit of $A_f$ when wind velocity is huge in front of sound velocity $c_s$ as follows:
\begin{align}
& &A_{f_0}&=A_f(V_w \gg c_s)=A \mathrm{exp}\left( \frac{V_0^2 }{2 }\cdot \frac{B}{A} \right).
\end{align}

The evolution of $A_f$ as a function of wind velocity can be retrieved in \cite{1987BAICz..38..222C}. Finally, we performed several computations using the Monte Carlo method to take into account the measurement errors of all the initial parameters  to obtain a ground map (strewn field) as a function of the final mass of the  bolide.

Of course, owing to the various simplifying assumptions made, we can only underestimate the size of the strewn field.  However, we can see that varying unknowns, such as the object density or the drag parameter, only cause the strewn field to slide along its center line. In the end, the main unknown is the width of that strewn field in the direction perpendicular to its center line, which can be several hundred meters up to 1~km.

Taking the example of the 1 January 2020 fall in Italy (\citealt{2020..Capodannomete}, \citealt{2020...Cavezzo}), our determined strewn field with a $99\%$ confidence level consisted of a thin strip $5.6$ km long and $100$ meters wide. The actual meteorite was found only $200$ meters from the central line of this strip. This demonstrates the accuracy of our method, the offset being mainly due to our approximating the meteorite shape as a sphere. 


\section{First results}
\label{sec:First-result}

\subsection{Statistics and network efficiency} 

One of the main objectives of the FRIPON network is to measure the unbiased incoming flux of extraterrestrial matter. In this section, we first present the raw statistics of detected falls. Next, we attempt to constrain the absolute flux of incoming material.
\begin{center}
\begin{figure}[!htbp]
\includegraphics[width=0.9\hsize]{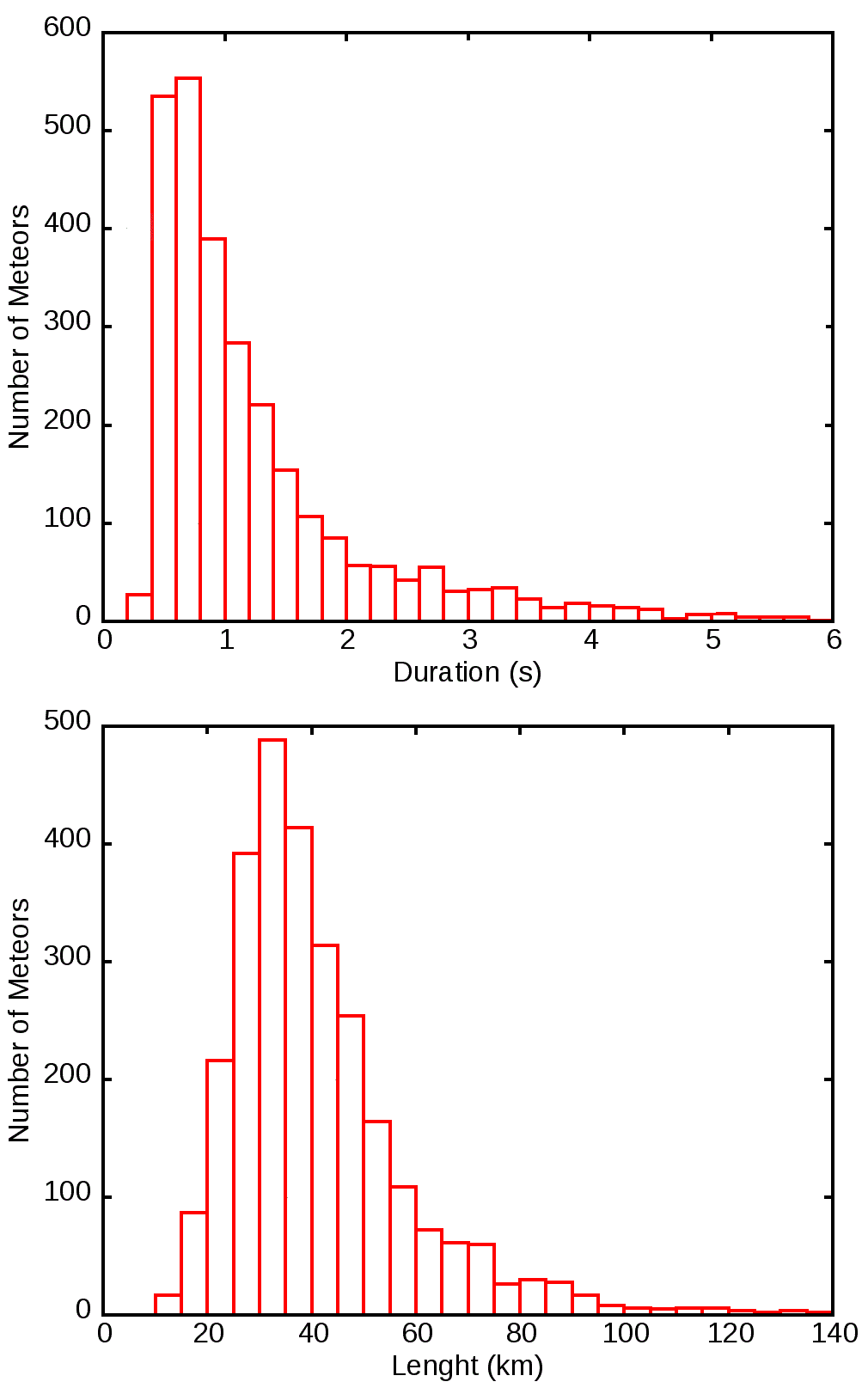}
\caption{Distribution of the duration (top) and length (bottom) of 3,200 bright flights. The cutoff for short exposures (less than 0.5s) is due to the acquisition software FreeTure \citep{2014pim4.conf...39A}.}
\label{Fig:lenth-duration}
\end{figure}
\end{center}

\begin{figure*}[!]
\includegraphics[width=1\hsize]{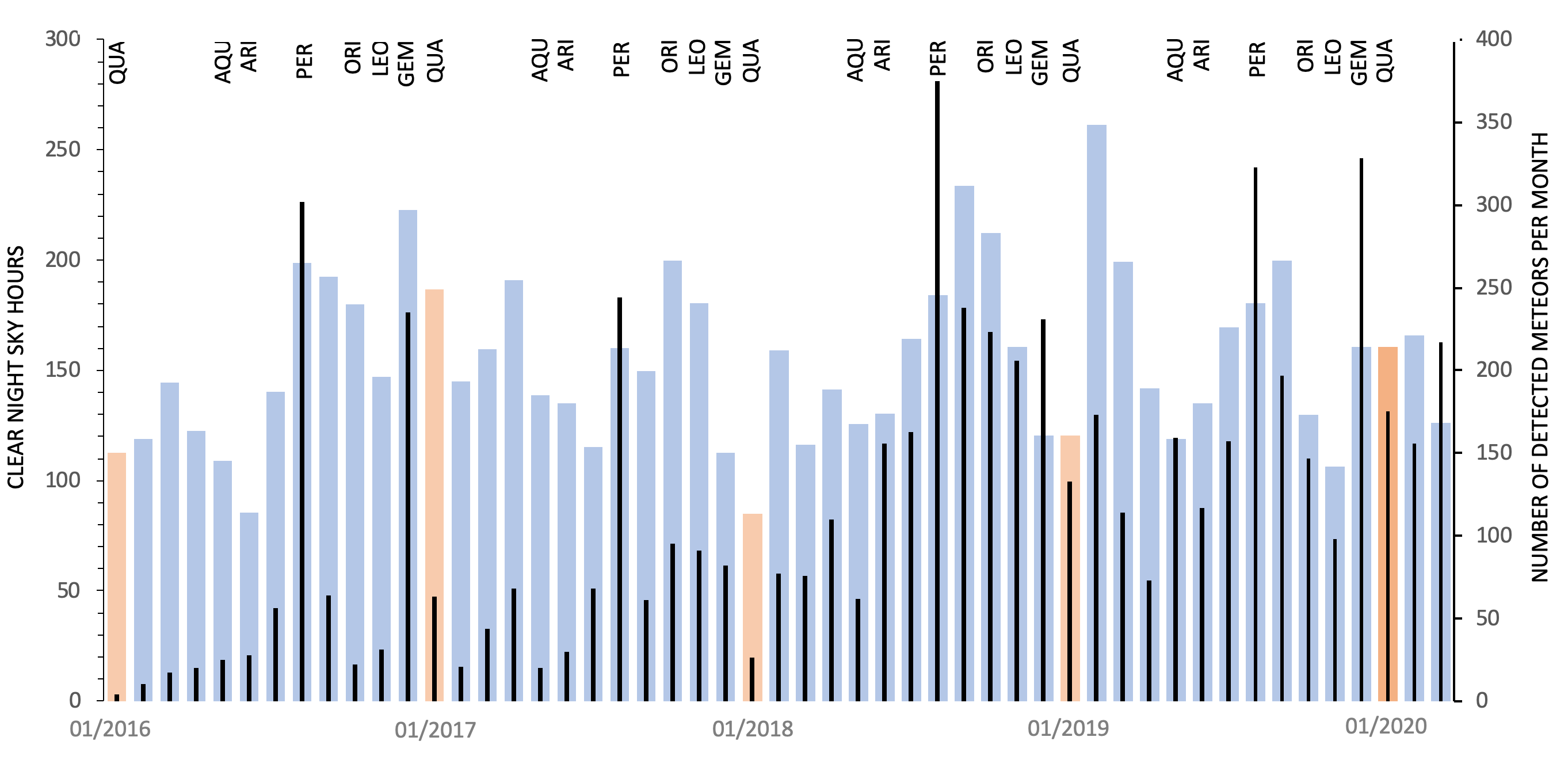}
\caption{Detection statistics for the last 3 years of operation. Of a total of 3,700 trajectories computed: double $58\%$, triple $20\%$, quadruple $8\%$, and more than 4 simultaneous detections $14\%$. The number of detections (black bars) gradually increases as the installation of the stations progresses. The blue bars (orange for January) indicate the number of clear night sky hours each month, making it possible to also visualize the effect of cloud cover. The main meteor showers are listed at the top.}
\label{Fig:HISTO-MUTIPLE}
\end{figure*}

\subsubsection{Raw meteoroid detections}

Fig.~\ref{Fig:lenth-duration} shows the histogram of duration and length of detected events. The average length of a meteor amounts to about $35$ km and it lasts for about 0.8s. Fig.~\ref{Fig:HISTO-MUTIPLE} shows the detection rate of the network between January 2016 and March 2020 as well as the average number of monthly clear night sky hours. Between 2016 and January 2019, we observed an increase in the number of detections that reflects the increasing number of installed cameras. Since January 2017, the annual number of detections appears to be fairly constant at around 1,000 detections per year.
Notably, the Perseid shower is the only shower standing out with regularity because of its high zenithal hourly rate (ZHR) and long duration. The shorter Geminid shower is less prominent (e.g., 2017 and 2018) because of greater cloud coverage. Weak meteor showers are not unambiguously detected in our data as a consequence of the photometric detection limitation of our cameras. As expected, our study shows a strong correlation between the monthly detection rate and the percentage of clear sky due to the local climate or to seasonal variations (or both, see Fig.~\ref{Fig:HISTO-MUTIPLE} and Fig.~\ref{Fig:FRIPONHeatMap} in Appendix). 

Fig.~\ref{Fig:RADIANT} shows the radiants of $3,200$ fireballs detected since 2016 and Table~\ref{tab:pluie_nombre} gives the number of detections for each shower per year. This figure presents that the main showers are detected and that the sporadic meteors are uniformly distributed over the celestial sphere except for part of the southern hemisphere, which is not at present within the reach of FRIPON. Overall, sporadic meteors represent $55\%$ of the data.

\begin{table}[ht!]
\caption{Number of meteors observed for the different meteor showers per year. The empty columns correspond to showers that fall outside the observation period from December $2016$ to December $2019$. The Quadrantides (QUA) were not observed in $2018$ owing to a power outage during the first half of January.}


\label{tab:pluie_nombre}
\centering
\begin{tabular}{c c c c c c}
\hline
\hline
Code & total & $2016$ & $2017$ & $2018$ & $2019$ \\
\hline 
GEM & $329$ & $86$ & $42$ & $82$ & $119$ \\
PER & $462$ & $-$ & $134$ & $174$ & $154$ \\
CAP & $38$ & $-$ & $4$ & $19$ & $15$ \\
QUA & $37$ & $-$ & $9$ & $-$ & $28$ \\
LYR & $27$ & $-$ & $13$ & $8$ & $6$ \\
LEO & $29$ & $-$ & $12$ & $16$ & $1$ \\ 
SDA & $37$ & $-$ & $9$ & $20$ & $8$ \\ 
ORI & $15$ & $-$ & $8$ & $7$ & $0$ \\ 
NTA & $33$ & $2$ & $11$ & $11$ & $9$ \\ 
MON & $11$ & $3$ & $2$ & $4$ & $2$ \\ 
SPE & $11$ & $-$ & $3$ & $5$ & $3$ \\ 
STA & $9$ & $-$ & $1$ & $6$ & $2$ \\ 
ETA & $5$ & $-$ & $1$ & $2$ & $3$ \\ 
HYD & $24$ & $6$ & $1$ & $8$ & $9$ \\
EVI & $12$ & $-$ & $7$ & $4$ & $1$ \\
JXA & $5$ & $-$ & $2$ & $3$ & $0$ \\
\hline
\end{tabular}
\end{table}

\begin{figure*}[!htbp]
\includegraphics[width=1\hsize]{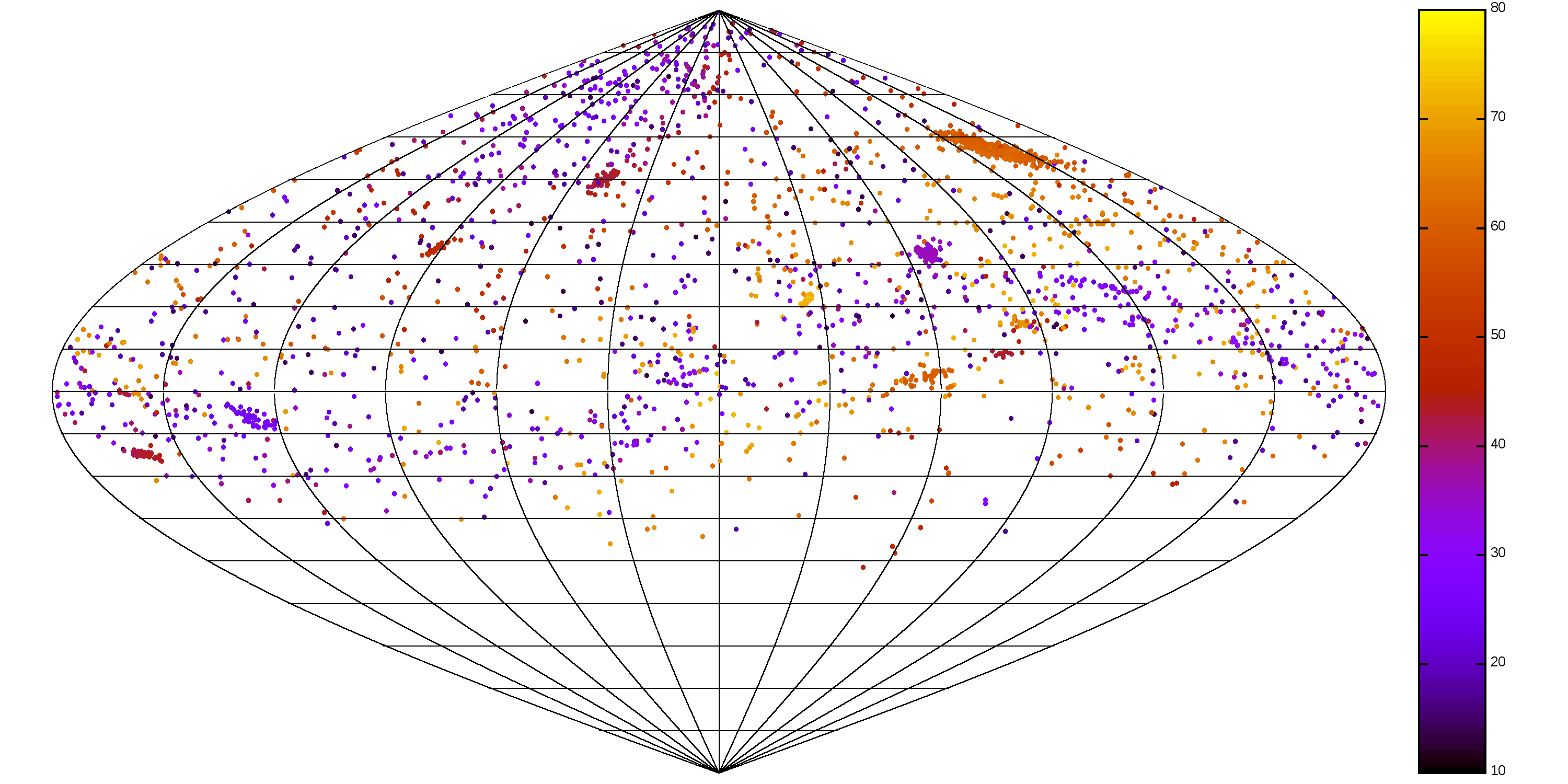}
\caption{ Fireball radiant in  Sanson-Flamsteed projection of equatorial coordinates from January 2016 to December 2019. The color scale corresponds to the initial velocity of the objects: 1) low velocities (in blue) for asteroidal like objects, 2) high velocities (in yellow) for cometary-like objects.  The main showers are detected. Of the objects, $55\%$ are sporadic: their radiants cover the sky uniformly except for its southern part, which is invisible from European latitudes.  The north toroidal sporadic source is visible in the top left corner and low speed objects are shown along the ecliptic plane coming from the anti-helion source.  }
\label{Fig:RADIANT}
\end{figure*}

\begin{figure}[!htbp]
\includegraphics[width=1\hsize]{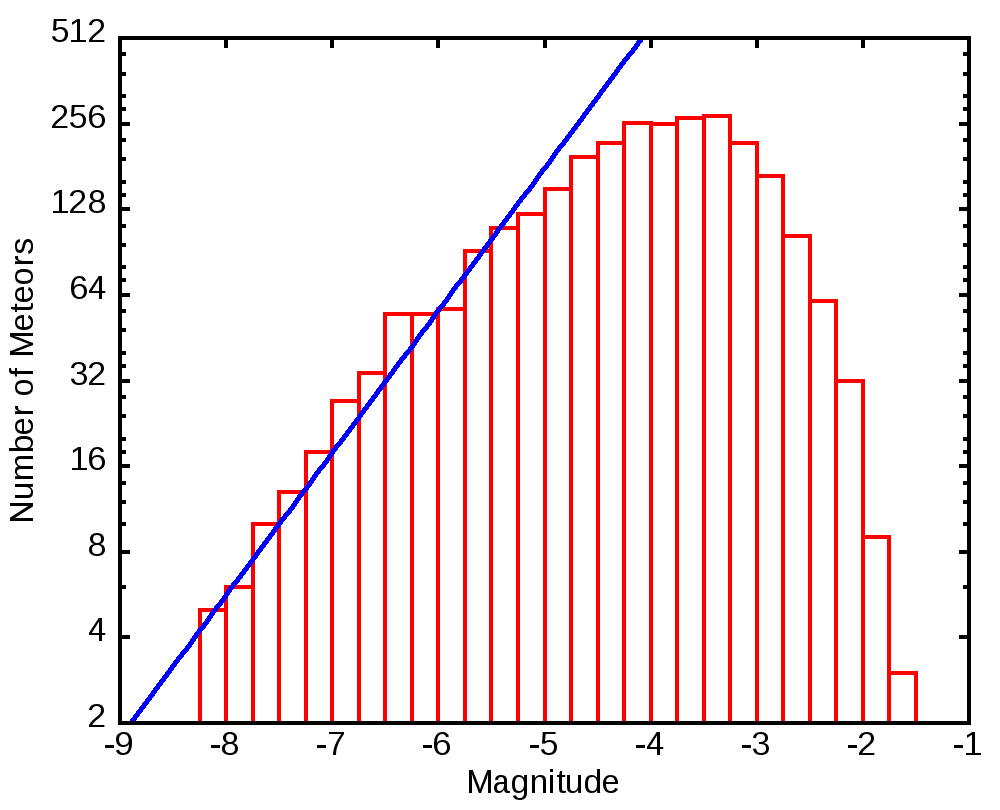}
\caption{Histogram of the absolute magnitude of all the events detected by the network showing that the exhaustive detection regime is only reached around mag $-5$. The slope is compatible with that obtained by previous studies such as \citet{2002Natur.420..294B}, as shown on Fig.~\ref{Fig:FRIPONSituation}, which describes the distribution of interplanetary matter from 1 cm to 1 km. The global shape of the histogram is similar to that in \cite{2014pim4.conf...23O}, which is shifted as CILBO cameras are more sensitive than FRIPON cameras.}
\label{Fig:PHOT-HISTO}
\end{figure}

\subsubsection{Quantifying the absolute meteoroid flux} 
\label{subsec:Quantifying}

An important goal of the FRIPON network is to estimate the absolute flux of incoming meteoroids. For this purpose, it is mandatory to measure the efficiency of the network in terms of meteoroid discovery. 

To estimate that flux, we need an estimation of the cloud coverage, the percentage of operational stations, and the sensitivity of our network as a function of meteor brightness. Regarding that last point, Fig.~\ref{Fig:PHOT-HISTO} shows the absolute magnitude histogram after three years of observations. Assuming a power-law size distribution for interplanetary matter \citep{2002Natur.420..294B}, it appears that FRIPON is clearly not fully efficient for events fainter than -5 in magnitude. This detection threshold is similar to that of the Prairie network  \citep{1996M&PS...31..185H} and implies, as for other networks \citep{2020arXiv200401069D}, a minimum detection size of $\sim$1 cm for incoming meteoroids. We note that smaller objects can nevertheless be detected if their entry speed is high enough. 

To calculate the efficiency of FRIPON, we only used the French stations as these were the first to be installed and France was fully covered in 2017. We considered its area, with a 120~km band added around it (Fig.~\ref{Fig:FRIPONHeatMap} in Appendix) for a total of $10^6$ km$^2$, which was the basis for the calculation. For $\ge$1 cm meteoroids (i.e., for  magnitude < -5 fireballs), we obtained an average rate of 250 events/year/$10^6$ km$^2$. Last, to estimate the incoming meteoroid flux for $\ge$1 cm bodies, we needed to correct for dead time (day time: 0.5 and average cloud cover: 0.4). The dead time-corrected meteoroid flux for $\ge$1 cm meteoroids is 1,250/year/$10^6$ km$^2$, which is comparable to the 1,500/year/$10^6$ km$^2$ value given by \cite{1996M&PS...31..185H}. Our determination is raw and requires that we carry out a more detailed analysis  in the future with more data. Our analysis shows that the network has reached a complete efficiency for the French territory for meteoroids larger than $1$ cm.

\subsubsection{Orbit precision} 

A precise determination of the orbit requires the extraction of a realistic initial velocity for the object. This can only be achieved by taking into account its deceleration in the upper atmosphere before the bright flight. Therefore our model of drag and ablation depends on three parameters (see section \ref{sec:drag-ablation}): the initial velocity $V$, a drag coefficient $A,$ and an ablation coefficient $B$.
Depending on the quality of the data---for example, the number of cameras, weather conditions, and distance of the camera to the bolide---these three parameters do not have the same influence on the trajectory calculation and cannot be determined with the same accuracy. 
We classified the meteors in three categories:

\begin{enumerate}
\item Those whose deceleration is hardly noticeable
($A/\sigma_A < 2$),
which represent $65\%$ of all meteors. 
\item Those for which only the deceleration is noticeable ($A/\sigma_A > 2$ and $B/\sigma_B < 2$), which represent $21\%$ of all meteors. In those cases, the ablation is not observed. 
\item  Those for which both the deceleration and the ablation are noticeable ($A/\sigma_A > 2$ and $B/\sigma_B > 2$), which represent $14\%$ of all meteors.
\end{enumerate}

For dynamical studies, only the detections that fall in one of the last two categories ($35\%$ of all detections) can be used. The typical velocity accuracy is then 100 m/s, which is required both for the identification of meteorite source regions in the solar system \citep{2018Icar..311..271G} and for the search for interstellar meteoroids \citep{2019msme.book..235H}.\\

\subsection{Dynamical properties of the observed meteoroids} 


\begin{center}
\begin{figure}[!htbp]
\includegraphics[width=1\hsize]{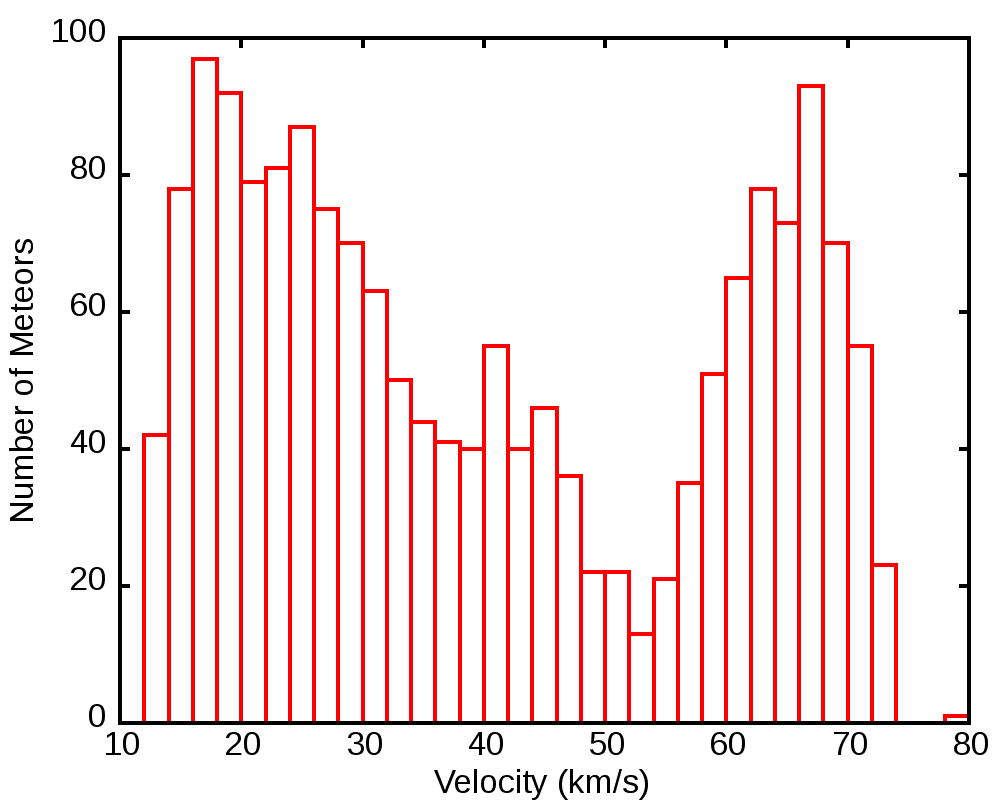}
\caption{Histogram of sporadic fireball entry velocities. Two populations can be observed: 1) low speed objects corresponding mostly to asteroidal orbits and 2) fast objects corresponding to TNOs or comet-like objects. This dichotomy has also been observed by \cite{2014pim4.conf...16D} with the CILBO network for smaller objects.}
\label{Fig:FRIPONTVitesses}
\end{figure}
\end{center}

\begin{center}
\begin{figure}[!htbp]
\includegraphics[width=1\hsize]{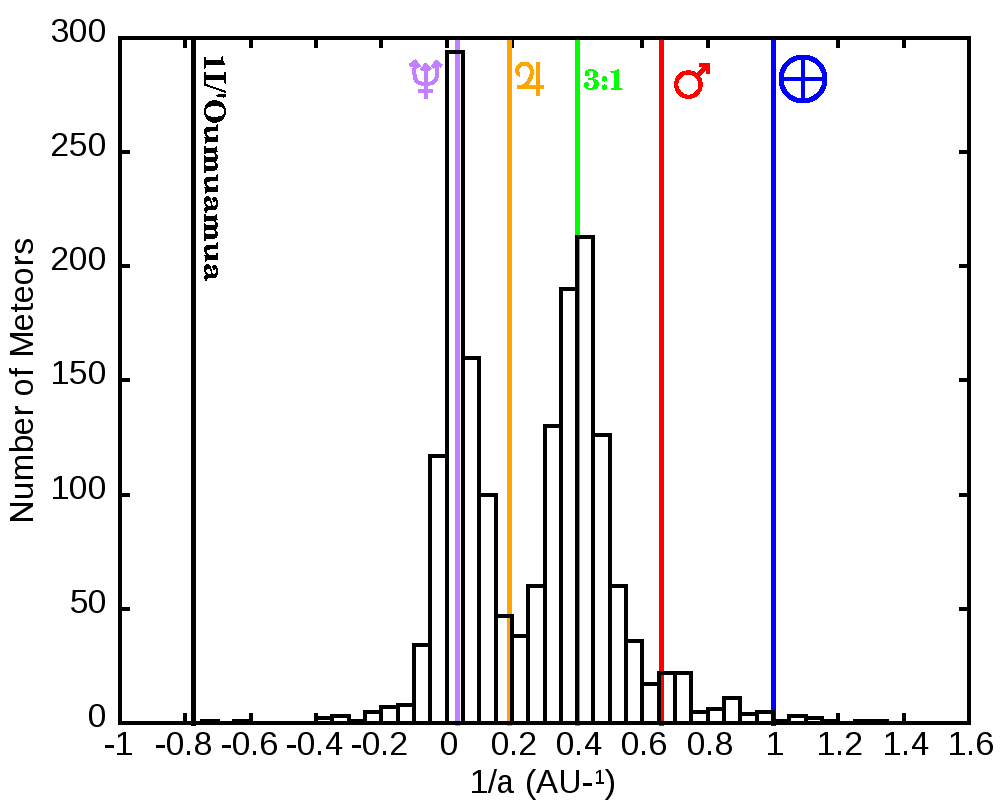}
\caption{Histogram of sporadic fireballs detected as a function of 1/a. This value is proportional to the orbital energy, making it possible to highlight two populations of objects: 1) the slow objects (of asteroidal origin) with a maximum related to the $3$:$1$ and $\nu 6$ resonances (green line), which are the main sources of NEOs; and 2) the fast objects around Neptune (purple line). These two populations are separated by Jupiter (orange line). The figure also shows the orbits of the Earth (blue), Mars (red), and the interstellar object 1I/Oumuamua (black). The FRIPON orbits with negative (1/a) values suffer from large errors and certainly do not correspond to orbits of interstellar objects. }
\label{Fig:FRIPONTResonances}
\end{figure}
\end{center}

In the following, we restrict our analysis to sporadic meteors. The histogram of initial velocities is shown in Fig.~\ref{Fig:FRIPONTVitesses}. It reveals two populations of meteoroids whose entry velocities differ by about 50~km/s, suggesting an asteroidal ($55\%$) and a cometary ($45\%$) population. This result can also be inferred from the histogram of meteoroid detections as a function of the inverse of the semimajor axis of their orbit (Fig.~\ref{Fig:FRIPONTResonances}). This figure clearly shows a main belt population with semimajor axes between that of Mars and that of Jupiter, as well as a cometary population, possibly including Oort cloud material, with semimajor axes greater than that of Jupiter. Last, we note the presence of a few meteoroids with negative semimajor axes. However, rather than concluding that interstellar matter was detected, we attribute these events to large errors associated with 
the calculation of their initial velocity. As a matter of fact, these events have semimajor axes that differ significantly from that of the interstellar object 1I/Oumuamua.

It is clear that in more than three years of observation, FRIPON has not detected any interstellar object so far. This compares to results obtained by other networks such as CMOR \citep{2004EM&P...95..221W}, who found that only 0.0008 $\%$ of the objects detected might be of interstellar origin; while a more recent work \citep{2018M&PS...53.1292M}  did not find interstellar candidate in CMOR data. In the case of the FRIPON network, only an upper limit of $0.1 \%$ can be given, but we expect the real value to be much lower.  
 \cite{2019msme.book..235H} showed that no network so far has ever experienced a conclusive detection of an interstellar meteoroid. Most false detections are likely to stem from a bad error estimation, especially that of the initial speed, which requires an estimation of the drag coefficient.

\subsection{Meteorite falls and first field search }  

\begin{table}[ht!] 
\caption{2016-2020  events with significant computed initial or final masses with a $m /\sigma_m > 2 $.  $\sigma_m$ is the standard deviation of the mass computed by the fit of our model. }
\centering
  \begin{tabular}{l c c c }
      \hline
      \hline
      Name & Date  & Initial  & Final  \\
           &       &   mass    &  mass  \\ 
           &       &    (kg)   &  (kg)     \\
      \hline
      Roanne    & 2016 08 06 & $1.6$ & $0.550$ \\
      Karlsruhe & 2016 09 25 & $5.3$ & $0.001$ \\
      Carlit    & 2016 11 27 & $3.0$ & $0.200$ \\
      Chambord  & 2017 03 27 & $1.0$ & $0.060$ \\
      Rovigo    & 2017 05 30 & $1.4$ & $0.150$ \\
      Golfe du Lion & 2017 06 16 & $12.2$  & $0.840$  \\
      Sarlat    & 2017-08-04 & $1.4$ & $0.110$ \\
      Avignon   & 2017 09 08 & $1.8$ & $0.005$ \\
      Luberon   & 2017 10 30 & $2.7$ & $0.017$ \\
      Menez-Hom & 2018 03 21 & $6.0$ & $0.001$ \\ 
      Quercy    & 2018 11 01 & $27.0$& $0.001$ \\
      Torino    & 2018 12 27 & $1.6$ & $0.550$ \\
      Sceautres & 2019 02 27 & $1.4$ & $0.110$ \\
      Gl\'enans & 2019 09 08 & $6.4$ & $0.540$ \\
      Saar      & 2019 10 13 & $1.3$ & $0.270$ \\
      Bühl      & 2019 10 16 & $1.2$ & $0.001$ \\
      Cavezzo   & 2020 01 01 & $9.1$ & $0.130$ \\
      Gendrey   & 2020 02 16 & $1.5$ & $1.100$ \\

      \hline
   \end{tabular}
   \label{table:recoveries}
\end{table}

Based on \cite{1989Metic..24..173H}, about ten meteorites weighing more than 100 g must fall each year over the area covered by the FRIPON network. Table~\ref{table:recoveries} lists the events that produced a computed significant initial and/or final mass. The fall rate that we observe for final masses equal or greater than 100 g is 2.7 per year. This value is compatible with that of \cite{1996M&PS...31..185H}, once corrected to take into account the $20\%$ overall efficiency of the FRIPON network (see above), as this yields a corrected  rate of 14 falls per year. Among these events, only 1 led to the recovery of meteorite fragments. This event occurred near Cavezzo in Italy \citep{2020..Capodannomete} and was detected by PRISMA cameras. Further details regarding the meteorite and its recovery will be presented in a forthcoming paper. This recovery is particularly important in showing that it is possible to find a 3 g stone thanks to the mobilization of the public with the help of various media (e.g., internet and newspapers). This strategy has worked well and can be reproduced for all comparatively small falls (typically a few dozen grams). In such cases, it is clear that the chances of finding the stone are low and do not warrant the organization of large searches, while an appeal to the general public may be fruitful. In the Cavezzo case, the meteorite was found on a path by a walker and his dog. 

It is also possible to calculate the meteorite flux for objects with final masses greater than 10 g and compare this value with previous estimates found by  \citealt{1989Metic..24..173H} (81/year/$10^6$ km$^2$), \citealt{1996MNRAS.283..551B} (225/year/$10^6$ km$^2$),
\citealt{2019Geo....47..673D} (222/year/$10^6$ km$^2$), and
\citealt{10.1130/G46733.1} (149/year/$10^6$ km$^2$).


We chose to compute the flux of objects with final masses greater than 100 g for which the accuracy is moderate to high ($m/\sigma_m > 2 $).  This flux is 14 meteorites/year/$10^6$ km$^2$ (see above). We extrapolated it down to a mass of 10~g, assuming a power-law distribution of the final masses of the meteorites \citep{1990Metic..25...41H}, and obtained a value of 94 meteorites/year/$10^6$ km$^2$, close to the value from \cite{1989Metic..24..173H}; this is also based on fireball data. This value is, however, lower than the other estimates (\citealt{1996MNRAS.283..551B} and \citealt{2019Geo....47..673D}), which are based on field searches. The \citealt{10.1130/G46733.1} estimate based on the study of meteorites found in Antarctic blue ice gives a mid-range value that is consistent with all previous estimates.

\section{Perspectives}   

\subsection{Extension of the network}

\begin{center}
\begin{figure}[!htbp]
\includegraphics[width=0.99\hsize]{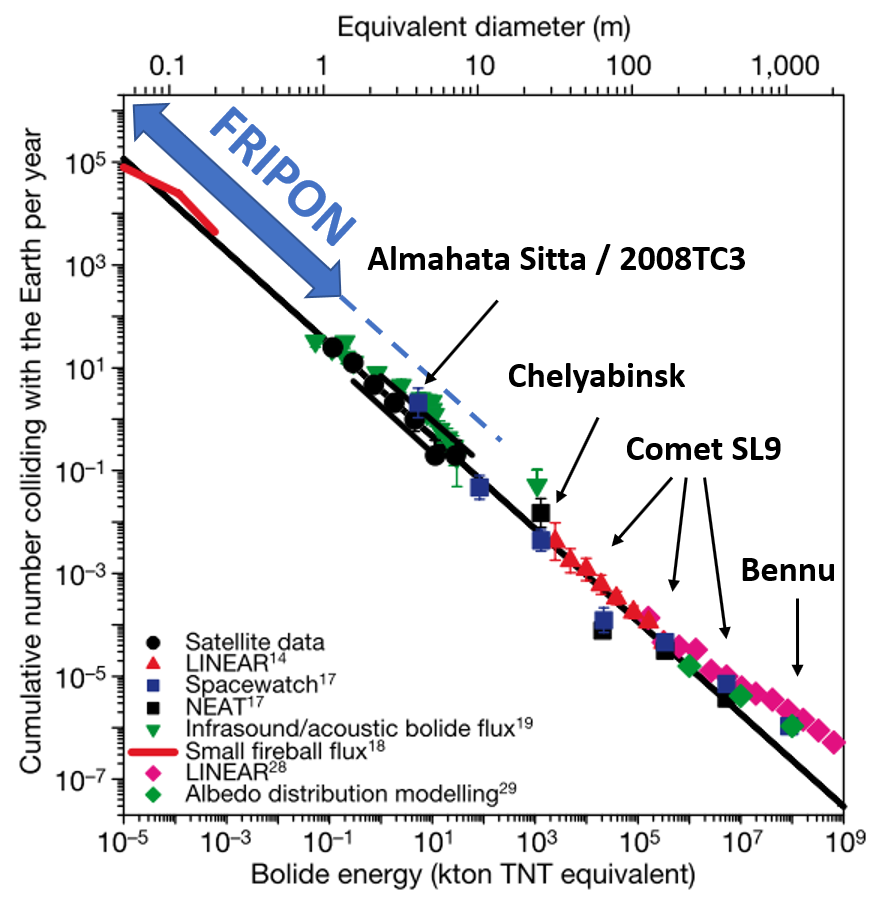}
\caption{Flux of small near-Earth objects colliding with the Earth \citep{2002Natur.420..294B}. Data are shown over a range of 14 magnitudes in energy. The statistical model is based on near-Earth population for big sizes and, for the smaller objects, it is derived from a decade-long survey of ground-based observations of meteor and fireballs. The FRIPON network lies exactly between minor planets (detected by telescopes and planetary impacts) and interplanetary dust (detected by meteor networks). The solid arrow corresponds to FRIPON nominal mode; the dashed line is for rare events, observable by FRIPON but with a very low probability.}
\label{Fig:FRIPONSituation}
\end{figure}
\end{center}

Significantly increasing the area covered by the network (by at least an order of magnitude) will be fundamental in increasing the recovery rate of meteorites, as this will lead to the detection, over a reasonable period, of a statistically significant number of very bright meteors that might be recovered on the ground as meteorites. Hence, there is a major interest in extending the FRIPON network over all of Europe and to other parts of the world. Such an extension has already begun (see Fig. \ref{Fig:NETWORK}) and will be pursued over the coming years. The development plan includes, as a priority, the densification of the European coverage as well as its extension to southern countries such as Morocco, Algeria, and Tunisia. 
For Spain, FRIPON is complemented by the SPMN network \citep{2004EM&P...95..553T}, with which we already collaborate for trans-national events and with whom we organized a search for a possible meteorite fall in January 2019.
Such a southern extension would be sufficient to generate a network area about ten times larger than that of metropolitan France.
In addition, the network is currently also being developed in Canada in North America and in Chile in South America. 
Fig.~\ref{Fig:FRIPONSituation} shows that 30 objects larger than one meter fall on Earth ($510$ $\times$ $10^6$ km$^{2}$) every year. Taking into account the current surface area of the FRIPON network, the average expected detection rate of such objects is limited to an average of one in ten years.
Extending the area of the network is thus a necessity to reach an acceptable detection rate for 1 meter objects. An extension to Europe and North Africa would make it reach a surface area of $6$ $\times$ $10^6$ km$^{2}$, which is comparable to that of the Australian DFN network \citep{2016pimo.conf...60D}, leading to a probability of a one-meter event approximately every years.

\subsection{Software}

The reduction pipeline is operational and only requires minor improvements. The acquisition software FreeTure still shows a surprisingly high false detection rate, which requires that daylight observations are turned off at the moment. A new version using deep learning techniques is being developed so that daytime observations will become possible. The development of a tool to compute the light curve of heavily saturated events \citep{2019EPSC...13.1758A} is also planned.

\subsection{Hardware}

The hardware currently in use in the network corresponds to pre-2014 technology.
A complete hardware update after five years of utilization is thus desirable to improve the temporal resolution of the light curves and the performance and flexibility of the acquisition computers. A non-exhaustive list of improvements includes upgrading from CCD to CMOS detectors and switching the current PCs to Raspberry Pi4 single board computers (SBCs).

\begin{figure}[!htbp]
\centering
\includegraphics[width=1.0\hsize]{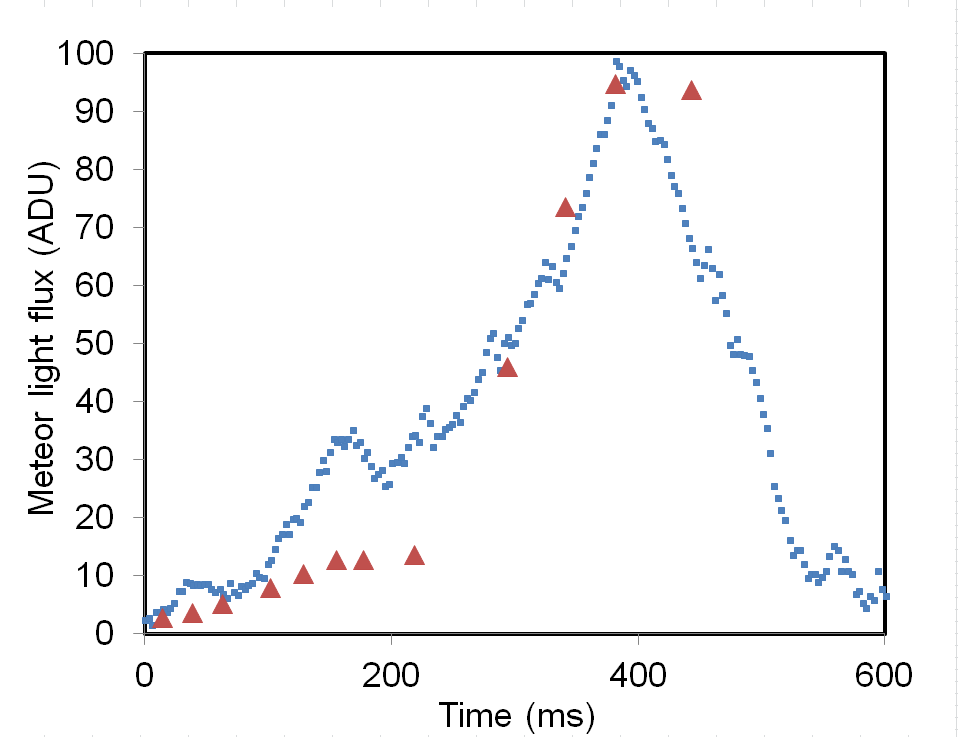}
\caption{Raw light flux from a bolide observed on 14 August 2019 at 03h07m02s UTC. Red triangles: Dijon FRIPON camera data (30 Hz, 12bits). Blue squares:  Radiometer prototype data (20 kHz, 14 bits).
The faster acquisition rate and the higher amplitude dynamic range of the radiometer allows more detailed observations of the meteor fragmentation and of high speed luminosity  variations.}
\label{Fig:Flux}
\end{figure}

In addition, a prototype of an all-sky radiometer is presently under development \citep{2019arXiv191104290R}, to resolve the saturation issue and improve on the bandwidth of the cameras. This radiometer covers the visible and near-infrared wavelengths. It is based on a 16 PIN photodiode matrix, followed by a trans-impedance amplification chain and a 14 bit industrial USB data acquisition module, which samples at a rate of 20 kHz. 
As an example, we superimposed on Fig.~\ref{Fig:Flux} the FRIPON camera light curve 
for an event of magnitude -9.5, which occurred on 14 August 2019 at 03:07:02 UTC and the corresponding high data rate radiometer light curve.

\subsection{Radio}

\begin{figure}[!htbp]
\includegraphics[width=1\hsize]{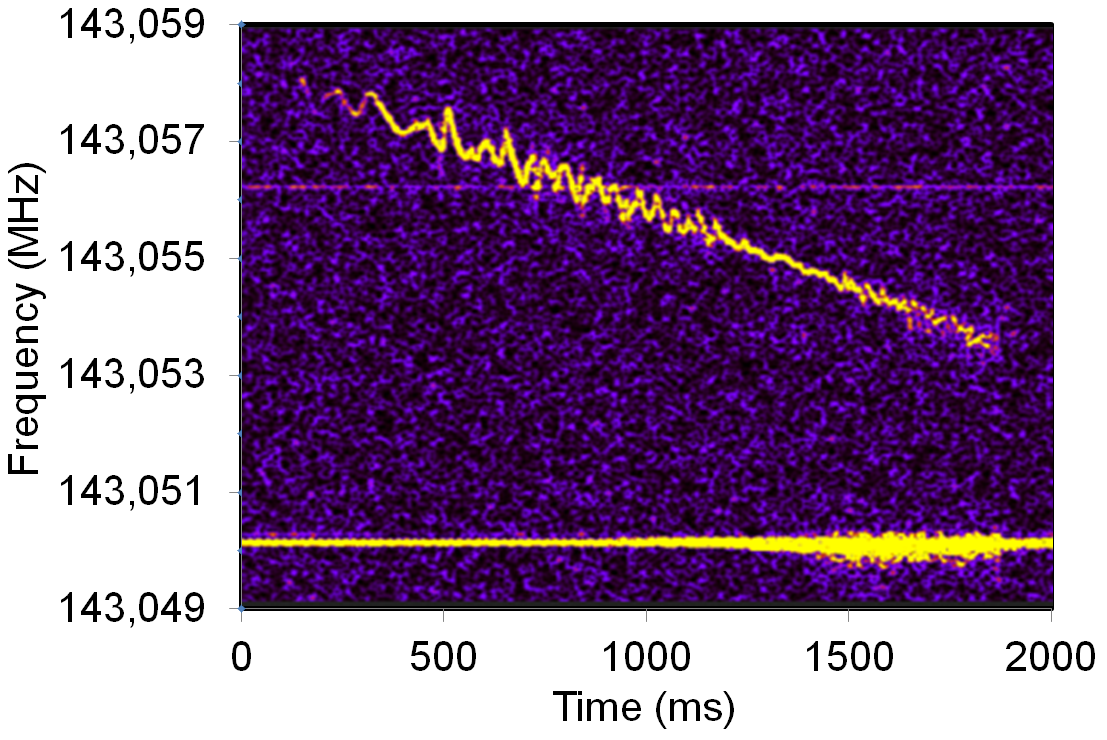}
\caption{Cyclic Doppler fluctuations on radio echo of the bolide observed on 8 August 2018 at 02h25m UTC, as seen by the Sutrieu radio receiver. Initial speed was $25.8$ km/s.}
\label{Fig:OSC}
\end{figure}

The aim of FRIPON radio receivers is an accurate measurement of meteor velocities through the Doppler effect, allowing a much better determination of the orbital data (especially semimajor axes). In Table~\ref{table:cameras}, we present the value of the initial velocity and effective surface-to-mass ratio derived for a meteor observed on 15 October 2018 at 1:15 UTC by five cameras. The accuracy achieved with the radio data leads to errors one order of magnitude lower compared to that achieved with only the visible images. However, it seems at present that only about 30 $\%$ of the optical detections lead to a detectable radio signal and that several bright radio events do not have any visible counterpart.
For this reason, radio data have not been widely used yet, and further work is needed to improve our understanding of the complex phenomena associated with the generation of radio echoes by the plasma surrounding the meteors. 
Over time, we came to the conclusion that detailed information on the fragmentation and final destruction of bolides 
might also be obtained thanks to the head echoes produced by the GRAVES HPLA radar.
Last, we sometimes detected unexpected oscillations on the usually smooth  Doppler shift curves  (\citealt{2018arXiv180405203R}), which indicates cyclic fluctuations on the radial positions of the radar cross section (RCS) of the plasma envelope surrounding the meteor bodies (see Fig.~\ref{Fig:OSC}).

\begin{table}[h] 
\caption{ 
For the 15 October 2018  event, first data reduction is based on all optical data. For the radio data, geometric model is first derived from the optical data. Then Doppler data are projected on the straight line of the trajectory, thus improving the speed and deceleration measurements by an order of magnitude.}
\centering
  \begin{tabular}{lcc}
      \hline
      \hline
      Sensor & Initial velocity & Effective surface / mass ratio \\
             &    $km/s$        &    $m^2/kg$   \\
      \hline
      Video & $66.49\pm 0.92$   &   <$1.28$      \\
      Radio+Video & $66.09\pm 0.09$   & $0.33\pm 0.14$  \\
      \hline
   \end{tabular}
   \label{table:cameras}
\end{table}

\subsection{Cross-reference data with infrasound network}

In recent years, infrasound has become an efficient technique, allowing for global detection of explosive sources in the atmosphere, and by extension of meteoroid atmospheric entries. There is an ongoing effort to improve the identification of valid signals and optimize the detection threshold for the International Monitoring System (IMS) developed to enforce the Comprehensive Nuclear-Test-Ban Treaty (CTBT) \citep{Marty2019}. Studies have determined that the IMS system, completed by experimental infrasound networks, is able to identify approximately 25$\%$ of fireballs with E > 100 t (TNT equivalent) energy and can provide key ground-based confirmation of the impact (timing and geo-location). This is particularly significant, as most impacts occur over the ocean, where no other instruments are likely to record the bolides \citep{2018imas.book..939S}.

It is expected that infrasonic observations of NEOs that regularly impact the Earth atmosphere will increase as the number of stations are deployed worldwide. Combining infrasound with optical observations, such as those collected by the dense network of cameras operated by FRIPON, would contribute to fill gaps in existing observation systems and help constraining source parameters, such as trajectory and energy deposition. The results become even more interesting in Europe, where the integration of national networks allows for a better characterization of smaller-energy events \citep{2019P&SS..17904715O}. The Atmospheric dynamics Research InfraStructure in Europe (ARISE) supports such multidisciplinary approaches by providing an extensive infrasound database for the estimation of NEOs potential risk and societal impact.

\subsection{Other records}

An extensive database of images covering a large area may be used for additional purposes. The study of transient luminous events (TLEs), such as sprites or spatial debris re-entries, may be cited as examples \citep{2018arXiv181112680C}. Since the Summer of 2017, the software FreeTure contains an experimental real-time algorithm for the detection of TLEs. This algorithm runs along with the meteor detection part on selected stations with a view to help localize TLEs observed by the future CNES space mission TARANIS \citep{2017SPIE10564E..04B}. The FRIPON network infrastructure can also be used to conduct large-scale light pollution monitoring campaigns using the all-sky calibration images collected over time \citep{2018JQSRT.209..212J}.

\subsection{Observation from space}

The network can also be extended vertically by combining space measurements with ground measurements. Space-borne observations have several advantages, such as providing a wide geographical coverage with one camera, longer recording times, and no weather constraints. The small satellite sector is evolving very quickly, opening up new opportunities for scientific missions \citep{2019AdSpR..64.1466M}.  In particular, relatively inexpensive missions make it possible to design swarms of satellites or even constellations dedicated to monitoring the Earth and therefore meteors. In this framework, a Universitary Cubesat demonstrator called Meteorix is under study \citep{Rambaux19}.  The Meteorix mission is dedicated to the observation and characterization of meteors and space debris entering the Earth's atmosphere. 
The orbit chosen for Meteorix is a low Earth sun-synchronous orbit at an altitude of 500 km.
Such configuration will make it possible to detect on average a sporadic meteoroid entry per day and about 20 meteors during a major meteor shower. The nominal mission lifetime is one year. Three-dimensional astrometry and photometry would become possible in case of a detection over the FRIPON network.

\section{Conclusion}

The FRIPON scientific network, originally developed to cover the French territory, is now a fully automated network monitoring fireballs above part of western Europe and a small fraction of Canada. As of today, it consists of 150 cameras and 25 radio receivers covering an area of about $1.5$ $\times$  $10^6$ km$^2$. The level of automation of the network is such that a recovery campaign can be triggered only a few hours after a meteorite reached the surface of the Earth.   

The FRIPON scientific project has been monitoring meteoroid entries in western Europe since 2016, thereby allowing the characterization of the dynamical and physical properties of nearly 4,000 meteoroids. It has thus allowed us to significantly enhance the statistics of orbital parameters of meteoroids, while also searching for possible interstellar meteoroids. The FRIPON observations show that the distribution of the orbits of incoming bolides appears bimodal, comprising a cometary population and a main belt population. Sporadic meteors amount to about 55\% of all meteoroids. In addition, we found no evidence for the presence of interstellar meteoroids in our sample. Overall, it appears that the range of sensitivity of the FRIPON network encompasses particles originating both from comets and asteroids. A first estimate of the absolute flux of meteoroids bigger than 1 cm amounts to 1,250 /year/$10^6$ km$^2$, which is a value compatible with previous reports. We also estimate the flux of meteorites heavier than 100g to 14/year/$10^6$ km$^2$, which is a value compatible with data from other fireball networks but lower than those obtained from collecting meteorites. Finally, the first meteorite has been recovered in Italy following observations by the PRISMA network, a component of the FRIPON network.\\

Further extension of the FRIPON network is under way. In the coming years, it will be extended to  North and West Africa as well as Canada and to the southern hemisphere in South America and South Africa. The goal is to reach a size large enough to allow the recovery of at least one fresh meteorite per year.
In addition to the geographical extension of the network, technical developments will be conducted to improve the photometry of saturated images. Moreover, we plan to implement new algorithms in the detection software, so that daytime observations become possible and useful. 
Finally, we plan to fully exploit the radio network, both to improve current orbits and to reach a better understanding of the physical mechanism of meteoroid entries.

\begin{acknowledgements}
\textbf{FRIPON} was initiated by funding from ANR (grant N.13-BS05-0009-03), carried by the \emph{Paris Observatory, Mus\'eum National d'Histoire Naturelle, Paris-Saclay University and Institut Pyth\'eas (LAM-CEREGE)}. \\
\textbf{Vigie-Ciel} was part of the
\emph{65 Millions d'Observateurs} project, carried by the \emph{Mus\'eum National d'Histoire Naturelle} and funded by the French \emph{Investissements d'Avenir} program.\\
FRIPON data are hosted and processed at Institut Pyth\'eas SIP (Service Informatique Pyth\'eas),
and a mirror is hosted at IMCCE (Institut de M\'ecanique Céleste et de Calcul des \'Eph\'em\'erides / Paris Observatory) with the help of IDOC\footnote{https://idoc.ias.u-psud.fr} (Integrated Data and Operation Center), supported by CNRS and CNES.\\
\textbf{PRISMA} is the Italian Network for Systematic surveillance of Meteors and Atmosphere.  It is a collaboration initiated and coordinated by the Italian National Institute for Astrophysics (INAF) that counts members among research institutes, associations and schools\footnote{http://www.prisma.inaf.it}.
  PRISMA was partially funded by 2016 and 2020 Research and Education grants from \emph{Fondazione Cassa di Risparmio di Torino} and by a 2016 grant from \emph{Fondazione Agostino De Mari} (Savona). \\
\textbf{FRIPON-Bilbao} is supported by a grant from \emph{Diputacion Foral Bizkaia (DFB/BFA)}.\\
\textbf{FRIPON-MOROI} was supported by a grant of the Romanian Ministery of Research and Innovation, CCCDI - UEFISCDI, project number PN-III-P1-1.2-PCCDI-2017-0226/16PCCDI/2018 , within PNCDI III.\\
\textbf{Rio de Janeiro} camera is hosted and partially maintained by MAST (Museum of Astronomy and Related Sciences)/MCTIC.\\
The Meteorix project acknowledges supports from labex ESEP (Exploration Spatiale des Environnements Plan\'etaires), DIM-ACAV+ R\'egion \^Ile-de-France, Janus CNES, IDEX Sorbonne Universit\'es and Sorbonne Universit\'e.\\
We thank Maria Gritsevich for comments which have been very helpful in improving the manuscript.

\end{acknowledgements}

%
%

\bibliographystyle{aa}
\bibliography{descfripon}

\begin{appendix} 
\label{Stations}
\section{Countries and observation stations involved in FRIPON}

\subsection{Algeria}
\begin{center}
\begin{tabular}{|l|c|l|c|}
\hline
Station & Long & Lat & Alt  \\
\hline
  Alger & 3.033126E & 36.797014N & 342 \\
  Djelfa & 2.575811E & 36.401415N & 773 \\
  Khenchela  & 7.191020E & 35.144201N & 1330\\
  Mostaganem &0.656112E & 36.037056N & 502 \\
\hline
\end{tabular}
\end{center}

\subsection{Australia}
\begin{center}
\begin{tabular}{|l|c|l|c|}
\hline
Station & Long & Lat & Alt  \\
\hline
  Perth & 115.894493E & 32.006304S & 30 \\
  Zadko & 115.712002E & 31.355793S & 50 \\
\hline
\end{tabular}
\end{center}

\subsection{Austria}
\begin{center}
\begin{tabular}{|l|c|l|c|}
\hline
Station & Long & Lat & Alt  \\
\hline
  Wien & 16.359753E & 48.20525N & 180 \\
\hline
\end{tabular}
\end{center}

\subsection{Belgium}
\begin{center}
\begin{tabular}{|l|c|l|c|}
\hline
Station & Long & Lat & Alt  \\
\hline
  Bruxelles & 4.357075E & 50.796727N & 114 \\
  Liège & 5.566677E & 50.582574N & 240 \\
\hline
\end{tabular}
\end{center}
\subsection{Brazil}
\begin{center}
\begin{tabular}{|l|c|l|c|}
\hline
Station & Long & Lat & Alt  \\
\hline
  Rio de Janeiro & 43.223311W & 22.8955612S & 50 \\
\hline
\end{tabular}
\end{center}
\subsection{DOME - Canada}
\begin{center}
\begin{tabular}{|l|c|l|c|}
\hline
Station & Long & Lat & Alt  \\
\hline
  Louiseville & 72.949033W & 46.249248N & 30 \\
  Montebello & 74.937772W & 45.660370N & 70 \\
  Montréal & 73.550401W & 45.560745N & 30 \\
  Mont Mégantic & 71.152584W & 45.455704N & 1110 \\
  Val David & 74.207167W & 46.030661N & 327 \\
  Val Saint François & 72.311258W & 45.493749N & 200 \\
\hline
\end{tabular}
\end{center}

\subsection{Chile}
\begin{center}
\begin{tabular}{|l|c|l|c|}
\hline
Station & Long & Lat & Alt  \\
\hline
  Baquedano & 69.845453W& 23.335221S & 1500 \\
  Cerro Paranal & 70.390400W & 24.615600S & 2518 \\
  Cerro Tololo & 70.806279W & 30.169071S & 2207 \\
  Chiu-Chiu & 68.650429W& 22.342471S  & 2525 \\
  La Silla & 70.732559W& 29.260110S & 2400 \\
  Maria Helena & 69.666780W & 22.346554S & 1155  \\
  Ollagüe & 68.253721W & 21.224131S & 3700 \\
  Peine & 68.068760W& 23.681256S  & 2450 \\
  San Pedro & 68.179340W & 22.953465S & 2408 \\
\hline
\end{tabular}
\end{center}
\subsection{Denmark}
\begin{center}
\begin{tabular}{|l|c|l|c|}
\hline
Station & Long & Lat & Alt  \\
\hline
  Sonderborg & 9.798961E & 54.908907N & 190 \\
\hline
\end{tabular}
\end{center}
\subsection{Vigie-Ciel - France}
\begin{center}
\begin{tabular}{|l|c|l|c|}
\hline
Station & Long & Lat & Alt  \\
\hline
  Aix en Provence & 5.333919E & 43.491334N & 184 \\  
  Ajaccio & 8.792768E & 41.878472N & 99 \\
  Amiens & 2.298872E & 49.898572N &39 \\
  Angers & 0.600625W & 47.482477N & 58 \\
  Angoulème & 0.164370E & 45.649047N & 100 \\
  Arette & 0.741999W & 42.974571N & 1687 \\
  Arras & 2.765306E & 50.287532N & 80 \\
  Aubenas & 4.390887E & 44.621016N & 315 \\
  Aubusson & 2.165551E & 45.955477N & 447 \\
  Aurillac & 2.431090E & 44.924888N & 690 \\
  Albi & 2.137611E & 43.918671N & 192 \\
  Bangor & 3.186704W & 47.313333N & 57 \\
  Barcelonette & 6.642280E & 44.389977N & 1162 \\
  Beaumont les Valence & 4.923750E & 44.883366N & 174 \\
  Belfort & 6.865081E & 47.640847N & 374 \\
  Besançon & 5.989410E & 47.246910N & 311 \\
  Biguglia & 9.479848E & 42.616786N & 8 \\
  Bizanet & 2.873811E & 43.163547N & 85 \\
  Brest & 4.504642W & 48.408671N & 66 \\
  Caen & 0.366897W & 49.192307N & 58 \\
  Cahors & 1.445918E & 44.455450N & 126 \\
  Cailhavel & 2.125917E & 43.161526N & 254 \\
  Cappelle la Grande & 2.366590E & 50.996056N & 12 \\
  Caussols & 6.924434E & 43.751762N & 1279 \\ 
  Cavarc & 0.644886E & 44.687615N & 113 \\
  Chalon sur Saône & 4.857151E & 46.776202N & 186 \\
  Chapelle aux Lys & 0.659221W & 46.628912N & 141 \\
  Charleville-Mézières & 4.720703E & 49.738458N & 187 \\
  Chatillon sur Seine & 4.577100E & 47.864833N & 222 \\
  Compiègnes & 2.801346E & 49.401338N & 48 \\
  Coulounieix & 0.706613E & 45.154948N & 208 \\
  Dax & 1.030458W & 43.693356N & 36 \\
  Dijon & 5.073255E & 47.312718N & 285 \\
  Epinal & 6.435744E & 48.185721N & 363 \\
  Glux en Glenne & 4.029504E & 46.957773N & 688 \\
  Gramat & 1.725729E & 44.745122N & 330 \\
  Grenoble & 5.761051E & 45.192599N & 230 \\
  Gretz-Armainvilliers & 2.742281E & 48.742632N & 112 \\
  Guzet & 1.300228E & 42.787823N & 1526 \\
  Hendaye & 1.749324W & 43.377440N & 87 \\
  Hochfelden & 7.567531E & 48.756330N & 191 \\
  Hyères & 6.112921E & 43.095433N & 240 \\ 
  La Chatre & 1.866338E & 46.529210N & 28  \\
  La Ferté Bernard & 0.647542E & 48.185502N & 95 \\
  Laval & 0.782894W & 48.081912N & 103 \\
  Le Bleymard & 3.737160E & 44.504370N & 1196 \\
  Le Mans & 0.163854E & 48.015681N & 109 \\
  Les Angles & 4.753658E & 43.961583N & 80 \\
  Les Makes & 55.410097E & 21.198890S & 990 \\
  Le Vaudoué & 2.522362E & 48.362668N & 80 \\ 
  Le Versoud & 5.851000E & 45.211726N & 224 \\
  Lille & 3.071544E & 50.614975N & 35 \\
  Ludiver & 1.727798W & 49.630735N & 180 \\
  Lyon & 4.866197E & 45.779935N & 190 \\
  Maido & 55.383012E & 21.079594S & 2160 \\
  Mantet & 2.306972E & 42.477420N & 1555 \\
  Marigny & 0.417403W & 46.197592N & 59 \\
  Marseille & 5.436376E & 43.343690N & 130 \\
  Maubeuge & 3.987223E & 50.277947N & 145 \\
  Mauroux & 0.819706E & 43.919035N & 225\\
  Migennes & 3.509820E & 47.968880N & 130 \\
\hline
\end{tabular}

\begin{tabular}{|l|c|l|c|}
\hline
Station & Long & Lat & Alt  \\
\hline
  Montpellier & 3.865524E & 43.632674N & 74 \\
  Moulins & 3.319005E & 46.559871N & 217 \\
  Nançay & 2.195688E & 47.367857N & 136 \\
  Nantes & 1.554742W & 47.238106N & 26 \\
  Onet le Chateau & 2.585813E & 44.364935N & 552 \\
  Orléans & 1.943693E & 47.836332N & 120 \\
  Orsay, GEOPS & 2.179331E & 48.706433N & 174 \\
  Osenbach & 7.206581E & 47.992670N & 471 \\
  Paris, MNHN & 2.357177E & 48.843075N & 55 \\
  Paris, Observatoire & 2.336725E & 48.836550N & 88 \\
  Pic de Bure & 10.335099E & 36.880495N & 2560 \\
  Pic du Midi & 0.142626E & 42.936362N & 2877 \\
  Pierres & 1.532769E & 48.579869N & 165 \\
  Pleumeur Bodou & 3.527085W & 48.783253N & 35 \\
  Poitiers & 0.380783E & 46.565784N & 130 \\
  Pontarlier & 6.351011E & 46.914613N & 834 \\
  Porto Vecchio & 9.271180E & 41.599753N & 22 \\
  Puy-de-Dome & 2.964573E & 45.772129N & 1465 \\
  Querqueville & 1.692611W & 49.665715N & 21 \\
  Reims & 4.067164E & 49.243267N & 137 \\
  Rennes & 1.674733W & 48.105705N & 100 \\
  Roanne & 4.036814E & 45.996456N & 360 \\
  Rochechouart & 0.819906E & 45.823100N & 250 \\
  Rouen & 1.100422E & 49.447464N & 50 \\
  Royan & 1.048922W & 45.639012N & 15 \\
  Sabres & 0.746172W & 44.149087N & 85 \\
  Saint Bonnet Elvert & 1.908838E & 45.165080N & 539 \\
  Saint Denis de Jouhet & 1.866338E & 46.52921N & 280 \\
  Saint Julien du Pinet & 4.054800E & 45.133304N & 961 \\
  Saint Lupicin & 5.792866E & 46.397709N & 590 \\
  Saint Michel (OHP) & 5.714722E & 43.933010N & 558 \\
  Saint Quentin & 3.293955E & 49.862943N & 120 \\
  Salon de Provence & 5.098180E & 43.642734N & 89 \\
  Sarralbe & 7.021394E & 48.982666N & 229 \\
  Strasbourg & 7.762862E & 48.579825N & 165 \\
  Sutrieu & 5.626334E & 45.915575N & 867 \\
  Talence & 0.59296W & 44.807851N & 48 \\
  Tauxigny-St-Bauld & 0.832971E & 47.223431N & 97 \\
  Toulouse & 1.479209E & 43.562164N & 151 \\ 
  Troyes & 4.064624E & 48.270024N & 132 \\
  Vains & 1.446219W & 48.663646N & 16 \\
  Valcourt & 4.911772E & 48.616524N & 141 \\
  VandoeuvreLesNancy & 6.155328E & 48.655893N & 373 \\
  Vannes & 2.810623W & 47.503369N & 58 \\
  Wimereux & 1.605850E & 50.762740N & 19 \\
\hline
\end{tabular}
\end{center}

\subsection{Germany}
\begin{center}
\begin{tabular}{|l|c|l|c|}
\hline
Station & Long & Lat & Alt  \\
\hline
  Conow &  11.325496E & 53.220087N & 68 \\
  Fürstenberg & 8.747344E & 51.516789N & 330 \\
  Haidmühle & 13.758000E & 48.823000N & 820 \\
  Hannover & 9.822995E & 52.405035N & 80 \\ 
  Ketzur & 12.631277E & 52.495000N & 144 \\
  Oldenburg & 8.165100E & 54.908907N & 123 \\ 
  Seysdorf & 11.720225E & 48.545182N & 460 \\ 
  Spiekeroog & 7.713935E & 53.773939N & 10 \\
  Stuttgart & 9.103641E & 48.750942N & 300 \\
  Weil-der-Stadt & 8.860460E & 48.751819N & 420 \\ 
\hline
\end{tabular}
\end{center}

\subsection{PRISMA - Italia}
\begin{center}
\begin{tabular}{|l|c|l|c|}
\hline
Station & Long & Lat & Alt  \\
\hline
  Agordo & 12.031320E & 46.284320N & 600 \\
  Alessandria & 8.618194E & 44.923830N & 107 \\
  Arcetri & 11.254372E & 43.750590N & 100 \\
  Asiago & 11.568190E & 45.849170N & 1365 \\
  Barolo & 7.943960E & 44.611070N & 315 \\
  Bedonia & 9.6324870E & 44.507693N & 550 \\
  Brembate di Sopra & 9.582623E & 45.718831N & 295 \\
  Camerino & 13.065680E & 43.130570N & 670 \\
  Capua & 14.175158E & 41.121389N & 30 \\
  Caserta & 14.332310E & 41.072620N & 14 \\
  Castellana Grotte & 17.147777E & 40.875611N & 312 \\
  Cecima & 9.078854E & 44.814460N & 670 \\
  Cuneo & 7.540082E & 44.384776N & 559 \\
  Felizzano & 8.437167E & 44.912736N & 122 \\
  Finale Ligure & 8.327450E & 44.178270N & 35 \\
  Genova & 8.936114E & 44.425473N & 310 \\
  Gorga & 13.636000E & 41.392100N & 810 \\
  Isnello & 14.021338E & 37.939684N & 580 \\
  Lecce &  18.111235E & 40.335278N & 23 \\
  Lignan & 7.4783333E & 45.789861N & 1678 \\
  Loiano & 11.331773E & 44.256571N & 787 \\
  Luserna San Giovanni & 7.258267E & 44.827685N & 571 \\
  Medicina & 11.644608E & 44.524383N & 35 \\
  Merate & 9.4286111E & 45.705833N & 345 \\
  Montelupo Fiorentino & 11.043198E & 43.755337N & 500 \\
  Monteromano & 11.635978E & 44.138456N & 765 \\
  Monte Sarchio & 14.645457E & 41.063718N & 298 \\
  Napoli & 14.255361E & 40.862528N & 102 \\
  Navacchio & 10.491633E & 43.683200N & 15 \\
  Padova & 11.868540E & 45.401945N & 64 \\
  Palermo & 13.299417E & 38.187283N & 35 \\
  Piacenza & 9.725030E & 45.035376N & 77 \\
  Pino Torinese & 17.764939E & 45.041240N & 620 \\
  Pontevaltellina & 9.981636E & 46.190379N& 1207 \\
  Reggio Calabria & 15.660189E & 38.119310N & 100 \\
  Roma & 12.485338E & 41.894802N & 52 \\
  Rovigo & 11.795048E & 45.081666N & 15 \\
  SanMarcello Pistoiese & 10.803850E & 44.064155N & 1000 \\
  Sardinia Radio Telescope & 9.130760E & 39.281950N & 100 \\
  Savignano & 12.392745E & 44.089660N  & 100 \\
  Scandiano & 10.657597E & 44.591002N & 153 \\
  Serra la Nave & 14.978864E & 37.691831N & 1725 \\
  Sormano & 9.2285806E & 45.883000N & 1131 \\
  Trento & 11.140785E & 46.065509N & 500 \\
  Tricase & 18.366199E & 39.923622N & 94 \\
  Triestre & 13.875086E & 45.642691N & 412 \\
  Vicenza & 11.534934E & 45.558383N & 39 \\
\hline
\end{tabular}
\end{center}
\subsection{Mexico}
\begin{center}
\begin{tabular}{|l|c|l|c|}
\hline
Station & Long & Lat & Alt  \\
\hline
  San-Pedro-Martir & 115.465753W & 31.045931N & 2830 \\
  Ensenada & 116.666651W & 31.869425N & 50 \\
\hline
\end{tabular}
\end{center}
\subsection{Morocco}
\begin{center}
\begin{tabular}{|l|c|l|c|}
\hline
Station & Long & Lat & Alt  \\
\hline
  Casablanca & 7.634891W & 33.596191N & 15 \\
  Oukaimeden & 7.866467W & 31.206160N & 2725 \\
  Ben-Guerir & 7.936012S & 32.218554N & 460 \\
\hline
\end{tabular}
\end{center}
\subsection{Netherlands}
\begin{center}
\begin{tabular}{|l|c|l|c|}
\hline
Station & Long & Lat & Alt  \\
\hline
  Denekamp & 6.965788E & 52.414965N & 27 \\
  Dwingeloo & 6.234525E & 52.484699N & 16 \\
  Groningen & 6.5256694E & 53.249458N & 21 \\
  Noordwijk & 4.418402E & 52.218752N & 25 \\
  Oostkapelle & 3.537670E & 51.571920N & 4 \\
\hline
\end{tabular}
\end{center}
\subsection{MOROI - Romania}
\begin{center}
\begin{tabular}{|l|c|l|c|}
\hline
Station & Long & Lat & Alt  \\
\hline
  B\^{a}rlad & 27.671676E & 46.230847N & 81 \\
  Berthelot & 22.889832E & 45.614765N & 400 \\
  Boc\c{s}a & 21.777756E & 45.384465N & 283 \\
  Bucure\c{s}ti & 26.096667E & 44.413333N & 81 \\
  Dej & 21.230793E & 45.738060N & 101 \\
  Feleac & 23.593715E & 46.710241N & 800 \\ 
  Gala\c{t}i & 28.031919E & 45.419133N & 81 \\
  M\u{a}d\^{a}rjac & 27.134554E & 47.045297N & 200 \\
  M\u{a}ri\c{s}el & 23.075184E & 46.660976N &1200 \\
  P\u{a}ule\c{s}ti & 25.978060E & 45.006917N & 242 \\
  Timi\c{s}oara & 21.230793E & 45.738060N & 101 \\
\hline
\end{tabular}
\end{center}

\subsection{Per\'u}
\begin{center}
\begin{tabular}{|l|c|l|c|}
\hline
Station & Long & Lat & Alt  \\
\hline
  Arequipa & 71.493272W & 16.465638S & 2400 \\
  Caral      & 77.520278W & 10.893611S &  350 \\
  Moquegua & 70.678491W & 16.828119S & 3300 \\
  Pisac-Cusco& 71.849639W & 13.422278S & 2972 \\
  Puno & 70.015600W & 15.824174S & 3830 \\
  Samaca     & 75.759028W & 14.568028S &  325 \\
  Santa Eulalia & 76.661667W & 11.897667S & 1036 \\
  Sicaya     & 75.296444W & 12.040167S & 3370 \\
  Tarma      & 75.683330W & 11.418250S & 3056 \\
\hline
\end{tabular}
\end{center}

\subsection{Sénégal}
\begin{center}
\begin{tabular}{|l|c|l|c|}
\hline
Station & Long & Lat & Alt  \\
\hline
  Dakar  & 17.479673W & 14.704672N & 15\\
  Thies & 16.962996W & 14.793530N & 20\\
  Saint-Louis  & 16.062019W & 16.423375N & 10\\ 
\hline
\end{tabular}
\end{center}

\subsection{South Africa}
\begin{center}
\begin{tabular}{|l|c|l|c|}
\hline
Station & Long & Lat & Alt  \\
\hline
  Cape Town  & 18.477390E & 33.934400S & 25\\
  Sutherland & 20.810676E & 32.379791S & 1800\\
  Cederberg  & 19.252677E & 32.499412S & 1000\\ 
\hline
\end{tabular}
\end{center}

\subsection{SPMN - Spain}
\begin{center}
\begin{tabular}{|l|c|l|c|}
\hline
Station & Long & Lat & Alt  \\
\hline
  Barcelona & 2.119061E & 41.391765N & 97 \\
  Bilbao & 2.948512W & 43.262257N & 60 \\
  Montsec & 0.736836E & 42.024865N & 820 \\
\hline
\end{tabular}
\end{center}
\subsection{Switzerland}
\begin{center}
\begin{tabular}{|l|c|l|c|}
\hline
Station & Long & Lat & Alt  \\
\hline
  Saint Luc & 7.612583E & 46.228347N & 2200 \\
  Vicques & 7.420632E & 47.351819N & 600 \\
\hline
\end{tabular}
\end{center}
\subsection{Tunisia}
\begin{center}
\begin{tabular}{|l|c|l|c|}
\hline
Station & Long & Lat & Alt  \\
\hline
  La Marsa & 10.335108E & 36.880492N & 20 \\
  Sousse & 10.611125E & 35.812668N & 15 \\
\hline
\end{tabular}
\end{center}
\subsection{SCAMP - UK}
\begin{center}
\begin{tabular}{|l|c|l|c|}
\hline
Station & Long & Lat & Alt  \\
\hline
  Armagh & 6.649632W & 54.352350N & 75 \\
  Canterbury & 1.072080E & 51.273500N & 21 \\
  Cardiff & 3.177870W & 51.486110N & 33 \\
  East Barnet & 0.169234W & 51.637359N & 87 \\
  Harwell & 1.315363W & 51.572744N & 90 \\
  Honiton & 3.184408W & 50.801832N  & 170 \\
  Manchester & 2.233606W & 53.474365N & 70 \\
\hline
\end{tabular}
\end{center}

\label{Radio setup}
\section{FRIPON radio hardware description}

FRIPON radio setup  \citep{2014pim4.conf..185R} is a multi-static radar consisting of 25 distant receivers and a  high power large aperture (HPLA) radar. Thanks to its omni-directional reception antenna, each single radio station is able to receive scattered GRAVES echoes from a meteor, from its ionized trail and/or from the plasma surrounding the meteor body.

\begin{figure}[!htbp]
\includegraphics[width=0.95\hsize]{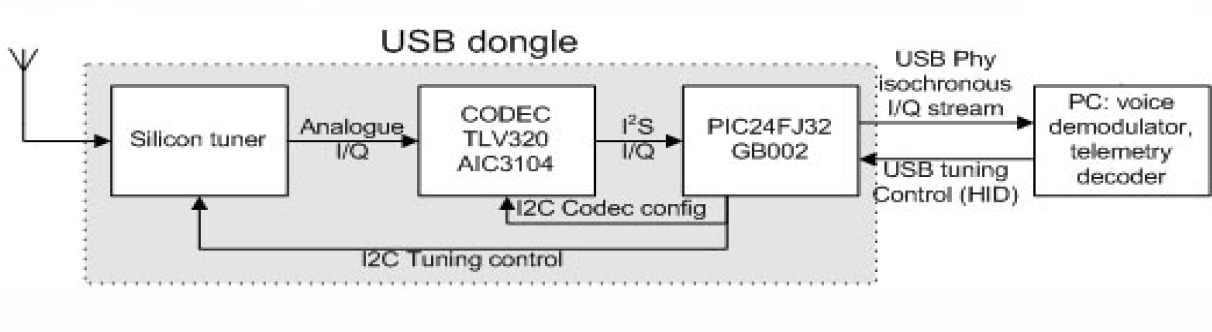}
\caption{Diagram of the FUNcube \citep{2013JBAA..123..169A} Software Defined Radio.}
\label{Fig:FUN}
\end{figure}

A typical FRIPON radio setup consists of
\begin{itemize}
\item  a 2.5 m long vertical ground-plane antenna ref. COMET GP-5N connected to the radio receiver via a $50 \Omega$ coaxial cable model KX4;
\item  a general purpose Software Defined Radio  (SDR) ref. FUNcube Dongle Pro + \citep{2013JBAA..123..169A}.
\end{itemize}
The ground-plane antenna radiation pattern is omni-directional in the horizontal plane, allowing both back and forward meteor scatter modes. The gain of this vertically polarized antenna is around  $6$ dBi.
The FUNcube SDR is connected to one of the USB ports of the station and the $I/Q$ data produced by the radio are recorded 24 hours a day on the local computer hard disk.
The SDR is a general coverage receiver (Fig.~\ref{Fig:FUN}), whose main characteristics are as follows:
\begin{itemize}
\item Frequency range $150$ kHz to $240$ MHz and $420$ MHz to $1.9$ GHz;
\item Sensitivity: typically $12$ dB SINAD NBFM for $0.15  \mu V$ at $145$ MHz;
\item Reference oscillator stability: 1.5 ppm; 
\item Sampling rate: $192$ kHz;
\item Bit depth: 16 bits (32 bits used internally).
\end{itemize}
A low noise amplifier (LNA) and a surface acoustic wave (SAW) filter fitted in the front end of each receiver offer an adequate sensitivity and selectivity for the meteor echoes.

\label{trajectory}
\section{Map of FRIPON meteor trajectories}

\begin{center}
\begin{figure*}[!]
\includegraphics[width=0.95\hsize]{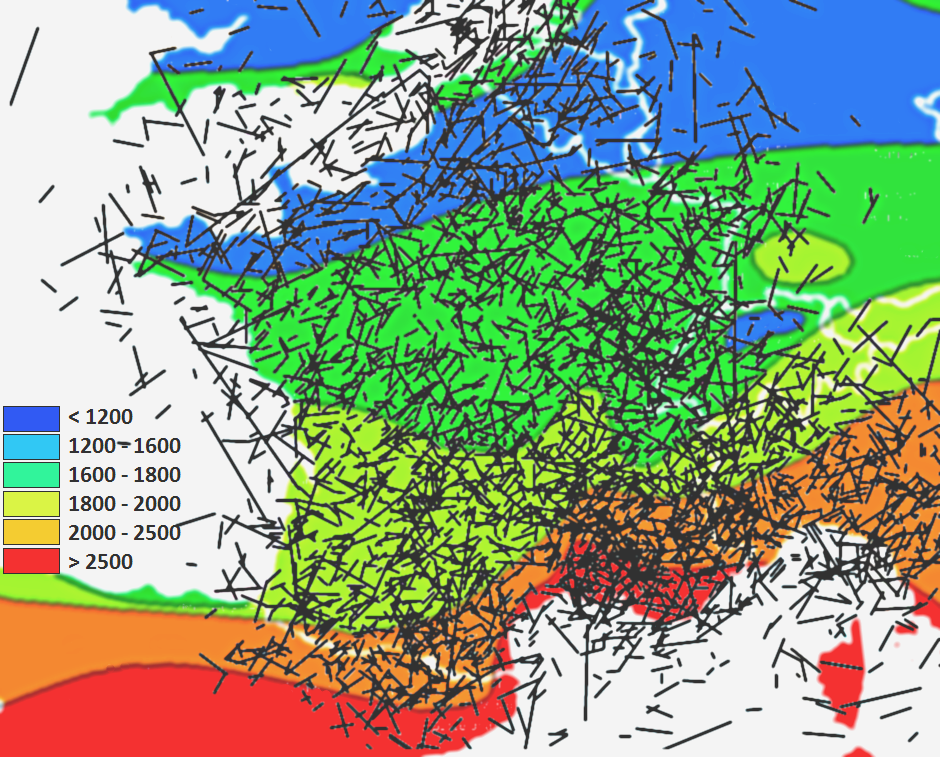}
\caption{Map of the 3,700 trajectories measured with FRIPON data from 2016 to early 2020. The concentration of detections is in part explained by the background sunshine weather map (sunshine duration in hours per year). The Rh\^{o}ne valley and the south of France have twice as many clear nights as the north. Another factor is that the installation of the cameras, done mostly throughout 2016, started in southern France and around Paris.}
\label{Fig:FRIPONHeatMap}
\end{figure*}
\end{center}

\end{appendix}

\end{document}